\documentclass[10pt,twocolumn,preprintnumbers,amsmath,amssymb,nofootinbib,superscriptaddress]{revtex4-1}

\usepackage{graphicx}
\usepackage{dcolumn}
\usepackage{amssymb,amsmath,bm}
\usepackage{color}
\usepackage[colorlinks,linkcolor=red,citecolor=blue,urlcolor=blue ]{hyperref}
\usepackage{multirow}
\usepackage{enumitem}
\usepackage{mathptmx} 
\usepackage[utf8]{inputenc}
\usepackage{tensor}
\usepackage{amssymb} 
\usepackage{lettrine}
\usepackage[normalem]{ulem}
\usepackage{empheq}
\usepackage{comment}


\begin{document}

  \title{Black hole spectroscopy with nonlinear quasi-normal modes}
  \author{Macarena Lagos}
  \email{macarena.lagos.u@unab.cl}
  \affiliation{Instituto de Astrof\'isica, Departamento de F\'isica y Astronom\'ia, Universidad Andr\'es Bello, Santiago, Chile}
  \author{Tom\'as Andrade}
  \email{tandrade@icc.ub.edu}
  \affiliation{Departament de F\'isica Qu\`antica i Astrof\'isica, Institut de Ci\`encies del Cosmos, Universitat de Barcelona, Barcelona, Spain} 
   \author{Jordi Rafecas-Ventosa}
  \email{jrafecasventosa@icc.ub.edu}
  \affiliation{Departament de F\'isica Qu\`antica i Astrof\'isica, Institut de Ci\`encies del Cosmos, Universitat de Barcelona, Barcelona, Spain} 
\author{Lam Hui}
  \email{lh399@columbia.edu}
  \affiliation{Department of Physics, Columbia University, New York, NY, USA} 
\begin{abstract}
The future detection of quasi-normal modes (QNMs) from black hole ringdown will allow for consistency and independent tests of general relativity (GR) in the strong-field regime.  
In this paper, we perform a ringdown Fisher forecast when including the dominant quadratic QNM (QQNM) expected in nearly equal-mass quasi-circular binary black holes (BBHs) observed by next-generation ground-based detectors, Einstein Telescope (ET) and Cosmic Explorer (CE). 
We consider a ringdown model with a total of four modes: three linear QNMs labeled by $(\ell m n)=(220), (330), (440)$ and one QQNM coming from the self-interaction of the dominant linear (220) QNM. We perform a forecast in two scenarios, when the QQNM parameters are considered to be: (a) independent of the linear QNMs; (b) dependent on the (220) QNM parameters, according to GR. 
In Scenario (a) we find the QQNM to generally  be measured with better precision than the (440) mode but worse than the (330) mode.
As shown in the past, high-spin BBHs tend to have higher relative QQNMs. Even in such cases, we only expect to confidently detect and resolve these four independent QNMs for nearby events, below redshift $z\sim 0.5$ in ET and CE for intermediate-mass BHs. 
In Scenario (b) we find the QQNM to be extremely useful for improving the precision on the (440) parameters, with negligible improvements on the (220) parameters. In this case, the (440) parameters are expected to be measured even better than those of the (330) QNM. As a result, we expect to confidently detect and resolve the three independent linear QNMs for events even at high redshifts, up to $z\sim 35$ in ET and CE for intermediate-mass BHs. Therefore, thanks to the inclusion of the QQNM, virtually all second-generation BBH events will provide excellent consistency tests of GR. 
\end{abstract}

  \maketitle

\section{Introduction}\label{sec:intro}

The emission of gravitational waves (GWs) from a remnant black hole (BH), formed after the merger of two compact objects, has been identified for decades as a clean way to perform tests of General Relativity (GR) \cite{1980ApJ...239..292D, Dreyer:2003bv, Cardoso:2016ryw}. This is because perturbed black holes emit GWs at characteristic frequencies, known as quasi-normal modes (QNMs), that only depend on the mass and spin of the BH, according to GR \cite{PhysRevD.2.2141,PhysRev.108.1063, Teukolsky:1973ha, Chandrasekhar:1975zza}. Such a GW signal is known as the ringdown emission, and by measuring some of these ringdown frequencies, one could straightforwardly perform  consistency tests of GR and tests for modified gravity (see reviews in \cite{Kokkotas:1999bd, Ferrari:2007dd, Berti:2009kk, Konoplya:2011qq}).

Formally, the ringdown emission is predicted from GR using linear perturbation theory around a final stationary remnant BH. At intermediate times\footnote{At early times (i.e.\ soon after the merger) the GW signal is dominated by a prompt response, while at late times the signal is dominated by polynomial tails \cite{PhysRevD.34.384, Andersson:1996cm, PhysRevLett.74.2414}.}, the corresponding GW signal is well modeled by a linear superposition of QNM frequencies $\omega_{(\ell m n)}$ with amplitudes $A_{(\ell m n)}$ starting at some fiducial time $t_0$, in geometric units ($c=G=1$):
\begin{equation}
    h_+ - ih_\times = \frac{M}{r}\sum_{\ell m n} A_{\ell m n}\; e^{-i\omega_{\ell m n }(t-t_{\rm 0})}\; {}_{-2}S_{\ell m n}(\iota,\beta; \chi\omega_{\ell m n}), \label{eq:qnms}
\end{equation}
where $h_{+,\times}$ are the two linear (and real) GW polarizations, $M$ is the total mass of the remnant BH, $r$ is the distance to the observer, and ${}_{-2}S_{\ell m n}(\iota,\beta; \chi\omega_{\ell m n})$ are the spin -2 spheroidal angular harmonics, which are functions of two periodic angles $\iota$ and $\beta$, as well as the BH dimensionless spin $\chi$ and the QNM frequencies. For low-enough values of the spin ($\chi \lesssim 0.7$) one can approximate the spheroidal by spherical harmonics, ${}_{-2}S_{\ell m n}(\iota,\beta; \chi\omega_{\ell m n})\approx {}_{-2}Y_{\ell m}(\iota,\beta)$, making only percent level errors \cite{Berti:2014fga} for the BHs considered in this paper.

In Eq.\ (\ref{eq:qnms}), there is a sum over all possible QNMs, which correspond to an infinite discrete spectrum of complex frequencies, $\omega=\omega_R+i\omega_I$, labeled by two angular harmonic numbers $(\ell, m)$  and one overtone integer number $n$. The real part $\omega_R$ determines the oscillation timescale of the modes, whereas the complex part $\omega_I<0$ determines their exponential damping timescale. These QNM frequencies depend only on the mass and spin of the remnant BH, and they have been calculated for different BH parameters and harmonic numbers \cite{Berti:2005ys, Berti:2009kk, Stein:2019mop} to use for spectroscopic tests of gravity.

To date, multiple studies have been performed on current data \cite{Capano:2021etf, Finch:2022ynt,Isi:2022mhy, Cotesta:2022pci, Siegel:2023lxl} and forecasts for ground and space-based GW detectors (e.g.\ \cite{Berti:2005ys, Ota:2019bzl, Bhagwat:2019dtm, Pitte:2024zbi}) using linear ringdown QNM frequencies. However, GR is a nonlinear theory and recent studies on second-order perturbation theory  have confirmed the presence of quadratic quasinormal modes (QQNM) in binary black hole (BBH) relativistic numerical simulations \cite{Mitman:2022qdl, Cheung:2022rbm, Ma:2022wpv,Redondo-Yuste:2023seq,Cheung:2023vki, Zhu:2024rej}, extending thus Eq.\ (\ref{eq:qnms}).

The physics of QQNMs results from the interactions of two linear, parent, QNMs. As a result, both the QQNM frequency and amplitude are fully determined by the parent modes. Regarding QQNM frequencies, they are given by \cite{Gleiser:1998rw, Gleiser:1995gx, Ioka:2007ak, Nakano:2007cj,Lagos:2022otp}
\begin{eqnarray}
    \omega_{(\ell m n)\times (\ell ' m' n')} = \omega_{\ell m n} + \omega_{\ell' m' n'},
\end{eqnarray}
where the linear parents modes are labeled by $(\ell m n)$ and $(\ell ' m' n')$. For the most common, nearly-equal mass quasi-circular binaries, the dominant linear QNMs are labeled by $(\ell=2, |m|=2, n=0)$ which will source the dominant QQNM with a frequency\footnote{There will also be a dominant QQNM frequency with $m<0$ given by $\omega_{(2-20)\times (2-20)}=2\omega_{2-20}$ but, as will be discussed later, it will be directly related to that with $m>0$ so we omit it here, yet it will indeed be included in our analysis. Furthermore, a third dominant QQNM is expected to have frequency $\omega_{(220)\times (2-20)}=\omega_{220}+\omega_{2-20}$ but this mode is non oscillatory for the binaries considered here, and is part of the memory effects of the GW signal \cite{Mitman:2024uss}, which will not be analyzed in this paper. A review on different contributions to the memory GW signal can be found in \cite{Favata:2010zu}.}
\begin{eqnarray}\label{Qfreq}
    \omega_{(220)\times (220)}=2\omega_{220}.
\end{eqnarray}
Regarding QQNM amplitudes, several recent works have attempted to obtain the quadratic-to-linear relationship using numerical and analytical methods, obtaining different, even conflicting, results \cite{Kehagias:2023ctr, Redondo-Yuste:2023seq,Cheung:2023vki, Zhu:2024rej, Ma:2024qcv}. However, recent studies \cite{Bourg:2024jme, Bucciotti:2024jrv} have shown that it depends on the amplitude of parity odd and even linear QNMs, with recent works \cite{Bourg:2024jme, Khera:2024yrk} showing that the previous conflicting results were due to different QNM parity choices. When the parent modes are labeled by $(\ell=2, |m|=2, n=0)$ and the binary is not precessing (and it is thus fully parity even), the sourced QQNM will have an amplitude \cite{Bucciotti:2024zyp,Khera:2024yrk, Bucciotti:2024jrv} 
\begin{eqnarray}\label{Qamp}
    A_{(220)\times (220)}=0.154 e^{-i0.068 } \; A_{220}^2
\end{eqnarray}
Note that Eq.\ (\ref{Qamp}) was calculated for a final non-spinning BH, yet most remnant BHs from mergers are expected to have final dimensionless spins over $\chi>0.6$, including the ones considered in this paper. While the 
quadratic form of Eq.\ (\ref{Qamp}) is always valid, the constant factors relating $A_{(220)\times (220)}$ and $A_{220}^2$ will depend on spin \cite{Khera:2024yrk}. We will still use the values in Eq.\ (\ref{Qamp}), since  they are expected to be modified only weakly with spin \cite{Cheung:2023vki, Redondo-Yuste:2023seq, Zhu:2024rej,Khera:2024yrk}\footnote{Based on the fittings in \cite{Cheung:2023vki}, an $18\%$ variation on the constant factor occurs when the spin changes from $\chi=0$ to $\chi=0.7$, if such factor evolves linearly with spin. A more precise derivation has been done in \cite{Khera:2024yrk}, where a difference of less than $10\%$ is observed between $\chi=0$ and $\chi=0.7$.}.

Importantly, while QQNMs contribute as sub-dominant modes in the total signal, the dominant QQNM expected in non-precessing nearly equal-mass binaries mentioned above has been shown to be larger than some linear QNMs previously studied in the literature \cite{Mitman:2022qdl, Cheung:2022rbm}. For this reason, a detailed analysis on the detectability of QQNMs is necessary. If detected
independently, such QQNM mode would allow for tests of the nonlinear behavior of GR around black holes. In this context, a recent study \cite{Yi:2024elj} obtained that the  next-generation  ground-based (XG) detectors, Einstein Telescope (ET) \cite{Branchesi:2023mws} and Cosmic Explorer (CE) \cite{Evans:2021gyd}, could detect QQNMs in up to a few tens of events per year. On the other hand, the future space-based detector, LISA \cite{Colpi:2024xhw}, could detect QQNMs in up to thousands of events during a four-year mission. A recent study analyzing the detectability of subdominant quadratic QNMs \cite{Khera:2024yrk} claims that several QQNMs could be observed by CE and LISA.

Given the promising future detectability of QQNMs, in this paper we perform a more detailed analysis for XG detectors, ET and CE. In particular, we will perform Fisher forecasts for quasi-circular  nearly equal-mass binary black holes (BBHs), representing a typical population of stellar-mass binaries given by first generation BHs (i.e.\ formed directly from stellar collapse), as well as intermediate-mass binaries likely given by second-generation BHs (i.e.\ formed by a previous BH merger). In these two type of binaries, not only the masses of the BHs but also the spins are expected to be different. Complementing the study in \cite{Yi:2024elj}, we will consider two possible scenarios where the QQNM plays an important role: (a) when the QQNM parameters are considered to be independent to the linear QNMs in the model, and its detection can thus be used to perform nonlinear tests of gravity; (b) when the QQNM parameters are considered to be dependent on its linear parent modes (based on GR), in which case its presence will help increase the confidence in the detection of the parent modes as well as other correlated linear QNMs, in order to perform more precise consistency spectroscopic tests of GR. 

In this paper, we first confirm previous results showing that second-generation BBHs are more promising for detecting QQNMs, due to their larger spins. Then, we perform a Fisher forecast for Scenario (a). We estimate the furthest distance at which ET and CE will be able to detect confidently and resolve independently three linear QNMs---namely (220), (330), and (440)---and one QQNM (coming from the self-interactions of the (220) linear mode). For second-generation BBHs, we find that this would only be possible for nearby events, below redshift $z \sim 0.5$ (and first-generation BBHs would have about ten times lower redshift reach). In such limiting redshift, we calculate the expected precision on the parameters of the four QNMs in the model, and provide scaling relationships to extend our results to different source masses and redshifts. 

Additionally, we perform a Fisher forecast for Scenario (b), providing again scaling relationships to estimate the future parameter precision for BBHs at different masses and redshifts. In this scenario, only the parameters for the three linear QNMs (220), (330), and (440) are fitted, with the parameters for QQNM $(220)\times(220)$ fixed by Eqs. (\ref{Qfreq}) and (\ref{Qamp}).
We find that the inclusion of the QQNM in the model helps improve by a factor of two the precision expected on the linear (440) parameters, while only marginally improving the parent (220) parameters. As a result of this improvement, the three linear QNMs can be detected confidently and resolved independently for second-generation BBHs even at high redshifts in ET and CE, up to $z\sim 35$ (and  the redshift reach is about fifty times lower for first-generation BBHs). Such results show the promise of including QQNMs in future ringdown models to perform precision tests of GR with virtually all detectable second-generation BBHs.  

This paper is organized as follows. In Sec.\ \ref{sec:QQNMs} we review the ringdown model used in this paper. In Sec.\ \ref{sec:fisher} we summarize the fisher forecast and  methodology used for assessing the detectability of QNMs. Then, we present the main results in Secs.\ \ref{sec:results1} and \ref{sec:results2} for Scenarios (a) and (b), with independent and dependent QQNM parameters, respectively. Finally, we provide discussions and conclusions in Sec.\ \ref{sec:conclusions}.

\section{Ringdown Signal}\label{sec:QQNMs}

In this section, we start reviewing the ringdown model used in this paper. For linear QNMs, we will use the model in Eq.\ (\ref{eq:qnms}). In practice, even though there are infinitely many QNMs, we will only include the dominant QNMs expected in typical quasi-circular nearly equal-mass black hole binary systems. In those cases, we include only the linear $(220)$, $(330)$ and $(440)$ modes. 
Indeed, most past forecasts analyzing linear QNMs have concentrated on these three modes as well as possible overtones of the $(22)$ angular harmonic (e.g.\ \cite{Bhagwat:2019dtm, Ota:2019bzl, JimenezForteza:2020cve, Baibhav:2020tma, Yi:2024elj, Pitte:2024zbi}). In order to keep the model simple, we do not include overtones. In fact, overtones have been shown to be sometimes harder to detect than higher angular harmonics, although that must be analyzed case by case, depending on BH spins and mass ratios \cite{JimenezForteza:2020cve, Baibhav:2023clw,Ota:2019bzl}. In any case, the results presented here for the QQNM are not expected to change if the (22) overtones are included.

As a result, at linear order we will consider:
\begin{align}\label{Lmodel}
    h_{+}-ih_{\times} \approx & \frac{M}{r}\left\{  A_{2 \pm2 0}\; e^{-i\omega_{2 \pm 2 0}(t-t_{\rm 0})}\; {}_{-2}Y_{2 \pm 2}(\iota,\beta)\right. \nonumber \\
    &+  A_{3 \pm3 0}\; e^{-i\omega_{3 \pm 3 0 }(t-t_{\rm 0})}\; {}_{-2}Y_{3 \pm 3}(\iota,\beta)\nonumber\\
    & \left.+ A_{4 \pm4 0}\; e^{-i\omega_{4 \pm 4 0 }(t-t_{\rm 0})}\; {}_{-2}Y_{4 \pm 4}(\iota,\beta) \right\}.
\end{align}
For the binaries studied here, the amplitudes and frequencies with signs $\pm$ will be related to each other. In particular, when binaries have small spins or large aligned spins, prograde (or co-rotating) QNMs are dominant \cite{Dhani:2020nik}, in which case we neglect retrograde QNMs and consider that $\omega_{\ell -|m| 0}=-\omega^{*}_{\ell |m| 0}$, where the real part of $\omega_{\ell |m| 0}$ is positive. Additionally, due to planar symmetry, the two $\pm$ modes will have amplitudes related by $A_{\ell -m 0 }= (-1)^\ell A^*_{\ell m 0 }$. Therefore, it is enough to analyze only the system for positive $m$ and straightforwardly predict the result for the negative $m$ modes.

In addition, in this paper we include one dominant quadratic QNM and analyze its parameter uncertainty as well as its impact on the parameter uncertainty of the rest of the ringdown modes.
Based on previous studies \cite{Mitman:2022qdl, Cheung:2022rbm, Cheung:2023vki}, we will only include the dominant expected QQNM, coming from the quadratic self-interaction of the linear $(\ell |m| 0)=(2 20)$ mode. In this case, we then add the following mode at quadratic order:
\begin{eqnarray}\label{Qmodel}
    h_{+}-ih_{\times}  \approx \frac{M}{r}\; A_{4\pm 4 Q}\; e^{-i\omega_{4\pm 4Q}(t-t_{\rm 0})}{}_{-2}Y_{4 \pm 4}(\iota,\beta),
\end{eqnarray}
where we have introduced a compact notation $A_{4 4Q}=A_{(2 20)\times (2 20)}$ and $\omega_{44Q}=\omega_{(220)\times (220)}$, given in Eqs.\ (\ref{Qamp})-(\ref{Qfreq}). Here, we can use the same symmetry properties mentioned above for the linear QNMs in order to relate the frequency and amplitude of the $(44Q)$ mode to those of the $(4-4Q)$ mode: $\omega_{4-4Q}=-\omega_{44Q}^*$ and $A_{4-4Q}=A_{44Q}^*$.

Note that in Eq.\ (\ref{Qmodel}) we have assumed that this dominant QQNM appears in the $(\ell m)=(4\pm 4)$ angular harmonics, which is expected to be exactly the case for non-spinning black holes, and approximately for low spinning black holes \cite{Lagos:2022otp}.

The final ringdown model to be analyzed will be given by both contributions from Eqs.\ (\ref{Lmodel})-(\ref{Qmodel}), assumed to be valid for $t>t_0$. While the value of the linear and quadratic frequencies are predicted from perturbation theory in GR (we use the \texttt{qnm} package \cite{Stein:2019mop} to obtain them), the value of the QNM amplitudes will be obtained by performing a time-domain fitting of the ringdown model to numerical relativity (NR) simulations. 

For the binaries considered here, we use the SXS simulation catalog \cite{Boyle:2019kee} to fit for the amplitude and phase of QNMs with fixed QNM frequencies with a least-squared fit. Given recent discussions on the validity of perturbation theory and the initial ringdown star time \cite{Baibhav:2023clw}, we perform the fittings at $t_0=10M$ after the merger time (identified as the time where the dominant $\ell=|m|=2$ angular harmonic amplitude peaks, with $M$ the mass of the remnant BH) \footnote{Even though we are not interested in detecting the overtone $(441)$ as it is expected to be subdominant, we will include it in the fitting procedure as it otherwise can bias the extracted amplitude of the QQNM $(44Q)$. In the later detectability analysis, the $(441)$ mode will not be considered.
}.

In Table \ref{table:QNM_fits} we collect the best-fit amplitudes of each QNM in the model (with positive $m$) at $t_0$, for two simulations SXS:BBH:0305 and SXS:BBH:0507. These two simulations are selected as representative BBH events from a population of first-generation BBH (i.e.\ low  spins), as well as a second-generation BBH (i.e.\ at least one large spin). In both cases, the mass ratio is close to unity, which is consistent with most binaries detected to date by the LVK collaboration \cite{KAGRA:2021vkt}. Moreover, in higher-mass ratio binaries the QQNM amplitude has been shown to be suppressed compared to the linear ones \cite{Mitman:2022qdl, Yi:2024elj} and thus such events are less promising and are not analyzed here. We note that the amplitudes and phases obtained in Table \ref{table:QNM_fits} do not exactly match Eq.\ (\ref{Qamp}). This is in part due to biases on the numerical procedures (e.g.\ fitting time\footnote{Indeed, the QQNM amplitude is known to suffer variations with fitting time, as shown in \cite{Mitman:2022qdl}. We still use $10M$ as it is the most common choice in the literature and makes comparisons to previous studies more straightforward.}, and usage of simulations not in the super-rest frame \cite{MaganaZertuche:2021syq}), and in part due to the fact that Eq.\ (\ref{Qamp}) was derived under some assumptions that do not exactly hold for our simulations (e.g.\ assumed non-spinning BHs and assumed that only the linear QNMs affect the QQNM amplitude but in general additional non-trivial time evolutions of the linear ringdown solution may also affect the QQNM amplitude). The fitting results obtained here are nonetheless in agreement with previous numerical studies \cite{Mitman:2022qdl}, and we will be using the values from Table I in the analyses below, except for Scenario (b).

\begin{table}[h!]
\centering %\vspace{-1em}
\begin{tabular}{|c|cc|} 
\hline
SXS ID\ & 0305 & 0507 \\
\hline
$q$ & $1.22$ &  $1.25$ \\\hline
$\chi_1$ & $0.33$ &  $0.8$ \\\hline
$\chi_2$ & $-0.44$ &  $0.4$ \\\hline
$\chi$ & $0.69$ &  $0.87$ \\\hline
$A_{220}$ & $4.2\times 10^{-1} e^{i 0.7}$ 
& $5.3\times 10^{-1}  e^{-i 0.4}$  \\\hline
$A_{330}$ & $2.3\times 10^{-2}  e^{i 0.4}$ 
& $2.3\times 10^{-2}  e^{i 0.7}$ \\\hline
$A_{440}$ & $2.2\times 10^{-2}  e^{-i 0.2}$ 
& $3.1\times 10^{-2}  e^{i 0.4}$  \\
$A_{44Q}$ & $4.1\times 10^{-2}  e^{i 2.5}$ 
& $6.2\times 10^{-2}  e^{i 0.4}$  \\
\hline
\end{tabular}
%\vspace{-1em}
\caption{\label{table:sims}%
List of simulations used (ID is shorthand for SXS:BBH:ID from the SXS catalog \cite{Boyle:2019kee}) with their mass ratios $q$, individual dimensionless spins  $\chi_{1,2}$, final remnant spin $\chi$, and best-fit amplitudes at $10M$ after the peak. Both binaries are non-precessing and are in quasi-circular orbits.}\label{table:QNM_fits}
\end{table}

While these two simulations can be used for any value of the binary masses, in the studies performed later on, we will assume that the BBH:0305 simulation is associated to stellar-mass BHs, while BBH:0507 to intermediate mass BHs.

We emphasize that, even though, the SXS simulations used here are not in the super-rest frame of the black hole \cite{MaganaZertuche:2021syq,Mitman:2022kwt}, for the simulation BBH:0305 we obtain percent level agreement in the amplitudes when comparing to the CCE simulations in the super-rest frame considered in \cite{Mitman:2022qdl}.

\subsection{Detected Ringdown}

In order to simulate realistic ringdown events, we need to account for cosmological propagation. This can be straightforwardly done by recovering physical units and incorporating the expanding universe. We thus replace: $r\rightarrow c^2d_L/G$; $\omega \rightarrow c^3\omega/G$, with $d_L(z)$ being the luminosity distance as a function of redshift; and $M\rightarrow M_z=M(1+z)$ being the redshifted mass of the remnant black hole. We use the Planck 2018 cosmological parameters \cite{Aghanim:2018eyx} to translate $z$ into $d_L$. Then, for any given black hole of mass $M$ and redshift $z$, we can calculate the observed amplitude and frequency of all QNMs.

Any given detector will have a response function, expressed as:
\begin{equation}
    h = F_+ (\theta,\phi,\Psi) \, h_+ + F_\times (\theta,\phi,\Psi) \,h_\times,
\end{equation}
where $F_{+,\times}$ are the responses to each GW polarization, which generally depend on the sky localization angles $(\theta,\phi)$ as well as the polarization angle $\Psi$. These three angles are independent of the $(\iota,\beta)$ angles in the ringdown spherical harmonics. The exact functional forms of $F_{+,\times}$ will depend on the shape of the GW detector. In this paper, we will only consider L-shaped detectors. Since we will be interested on representative ringdown events, we will only care about the average values of the detector responses, given by (e.g.\ \cite{Moore:2014lga}):
\begin{eqnarray}\label{Faverage}
      \langle F_+^2 \rangle = \langle F_\times^2 \rangle=\frac{1}{5}, \quad   \langle F_+F_\times \rangle =0. 
\end{eqnarray}
Similarly, we will only consider angular harmonic averages of the waveform:
\begin{equation}\label{Yaverage}
  \langle {}_{-2}Y_{\ell m}\; {}_{-2}Y^*_{\ell' m'}\rangle = \frac{1}{4\pi}\delta_{\ell \ell'}\delta_{m m'}, 
\end{equation}
where we are using the fact that angular harmonics are orthogonal and normalized to unity when integrated over the entire sphere.
    
For analyzing the detectability of QNMs, it will also be necessary to express the ringdown model in frequency domain. Throughout this paper, when transforming a ringdown signal from time domain to frequency domain, we will follow the approach outlined in \cite{Flanagan:1997sx}. In this case, in order to avoid high-frequency contamination from transforming a sharply-cut time-domain ringdown signal, we generate an effective time-domain ringdown signal by reflecting the original ringdown signal to $t<t_0$ and include a $1/\sqrt{2}$ factor to compensate for the double signal. Notice though that, based on the results in \cite{Berti:2005ys}, we do not expect such a reflection procedure to influence considerably the Fisher forecast results obtained later. 
Mathematically, we can write the ringdown signal in fourier-domain as:
\begin{equation}
    \tilde{h}(f) = \int_{-\infty}^{+\infty} dt e^{i 2\pi f t} \times \begin{cases}
        h(\omega, t)/\sqrt{2} & \text{if } t>t_0 \\
        h(\omega^{*},t)/\sqrt{2} & \text{if } t<t_0
    \end{cases}
\end{equation}
This integration can be performed analytically as shown in \cite{Berti:2005ys}, resulting in a Lorentzian function. 

Having a response function in frequency domain, one can then assess the detectability of QNMs by comparing to the sensitivity curve of a given GW detector. We will consider two cases: (i) Einstein Telescope as two aligned L-shaped detectors with 15km arm length, each one with a sensitive curve ET-D \cite{Hild:2010id}; (ii) Cosmic Explorer as one L-shaped detector with 40km arm length with the sensitive curve in \cite{CESn}.

\begin{figure}[h!]
\centering
\includegraphics[width=0.85\linewidth]{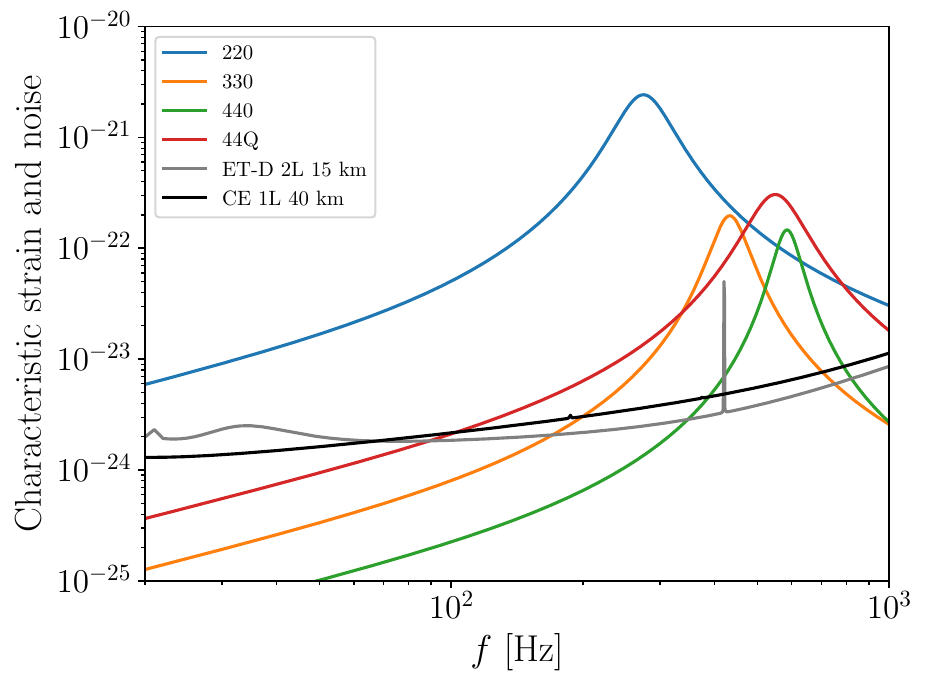}
\caption{\label{fig:ET}  Characteristic noise (black and gray lines)  for ET and CE detectors, compared to characteristic strain (colored lines) of the individual QNMs for a GW150914-like GW event.}
\end{figure}

The binary simulation SXS:BBH:0305 with a total mass $M=61M_\odot$ and redshift $z=0.093$ can be used as a proxy for the first BBH event detected by the LIGO-Virgo collaboration, GW150914 \cite{LIGOScientific:2016vbw}. In that case, we show in Fig.\ \ref{fig:ET} the dimensionless characteristic noise and strain of different QNMs, as function of frequency. The characteristic noise is defined as $S_c=\sqrt{f S_n(f)}$ with $S_n$ being the one-sided power spectral density \footnote{The sensitivity curve for ET was taken from \cite{ETSn}, which describes one L-shaped 10km long detector. Here, we have adjusted $S_n$ by a factor of $1/2$ for having two detectors, and an additional geometric factor of $4/9$ due to the 15km length change.}. The characteristic strain for a given QNM is calculated as $h_c=2f|\tilde{h}(f)|$ with $\tilde{h}$ being the frequency-domain ringdown waveform using the averaged detector response and harmonics in Eq.\ (\ref{Faverage})-(\ref{Yaverage}) (that is, replacing $F_{+,\times}\rightarrow 1/\sqrt{5}$ and ${}_{-2}Y_{\ell m} \rightarrow 1/\sqrt{4\pi}$ in the strain).

The shapes of the QNM curves in Fig.\ \ref{fig:ET} have a Lorentzian form, and the real part of the frequency $\omega_R$ determines where the curve peaks, whereas the imaginary part $\omega_I$ determines the width of the curve. We observe that the dominant mode is the (220), while the QQNM is comparable to the linear higher harmonics. 

\section{Methodology}\label{sec:fisher}
In this section, we summarize how to calculate the signal-to-noise ratio (SNR) for a given ringdown signal and provide details of the Fisher forecast procedure we will use to assess the detectability of QNMs and parameter precision. 

\subsection{Signal-to-Noise Ratio}
The SNR, $\rho$, for any given signal, is calculated as 
\begin{equation}
    \rho^2 = \langle h | h \rangle ,\label{eq:rho}
\end{equation}
where $h$ is the detector response, and this product is given by the frequency-domain noise-weighted overlap as:
\begin{equation}\label{Overlap_eqn}
    \langle h_1 | h_2\rangle = 2\int_0^\infty df \; \frac{\tilde{h}_1^*(f)\tilde{h}_2(f)+\tilde{h}_1(f)\tilde{h}_2^*(f)}{S_n(f)},
\end{equation}
with $S_n(f)$ being the one-sided noise power spectrum. Note this quantity is real and symmetric in $h_{1,2}$. Explicit expressions for the SNR in our ringdown model are provided in Appendix \ref{app:snr}. Note that, in general, the strain $h$ is a function of multiple parameters, but in Eq.\ (\ref{Overlap_eqn}) we only explicitly show its dependence on the frequency.

As previously discussed, in order to obtain results valid for averaged observer angular positions, we average over angles. Due to the orthogonality of spherical harmonics in Eq.\ (\ref{Yaverage}), one can calculate the SNR contribution of each angular harmonic $(\ell m)$, $\rho_{\ell m}$, and express the total SNR as:
\begin{equation}
    \rho^2= \sum_{\ell m} \; \rho^2_{\ell m}.
\end{equation}
In the case of linear QNM theory, if we consider the ringdown model in Eq.\ (\ref{Lmodel}), we will only have contributions from the fundamental modes, such that $\rho_{\ell m}=\rho_{\ell m 0}$. Thus, we can re-express the total SNR as
\begin{equation}
    \rho^2= 2\left[ \rho^2_{220} + \rho^2_{330}+\rho^2_{440}\right],
\end{equation}
where a factor of $2$ has been added to account for the contribution from negative $m$ modes. Nonetheless, when including the QQNM in Eq.\ (\ref{Qmodel}), a mixing will occur between the $(440)$ and $(44Q)$ modes \footnote{This is analogous to the mixing that would occur between the fundamental mode and higher overtones of a given angular harmonic, in the linear QNM theory.}.  In that case, the SNR associated to the $(\ell m)= (44)$ angular harmonic will be:
\begin{equation}
    \rho^2_{44}= \rho^2_{440}+ \rho^2_{44Q}+ \langle h_{440}| h_{44Q}\rangle, \label{eq:SNRmix}
\end{equation}
where the term $\langle h_{440}| h_{44Q}\rangle$ can be positive or negative depending on the relative phase evolution of the two modes. In this paper, we find the mixing term $|\langle h_{440}| h_{44Q}\rangle |$  to be comparable to $\rho^2_{440}$ and $\rho^2_{44Q}$, and it cannot be neglected.  
Based on SNR, one could assess the potential detectability of the QQNM $(44Q)$ by imposing its individual SNR to be over a given detectability threshold (this is the approach used in \cite{Yi:2024elj} and in previous works on linear QNM theory). We can define a conservative threshold of $\rho_{44Q}\geq 8$. In the mock event considered in Fig.\ \ref{fig:ET}, we obtain a QQNM SNR of $\rho_{44Q}=19$ for ET and  $\rho_{44Q}=13$ for CE. We would thus conclude that the QQNM could be detected with both ET and CE.

While this SNR condition is useful to assess the potential of observing the QQNM, we will confirm later that it does not ensure a confident detection of the QQNM and the rest of the ringdown modes together.  Imposing an SNR threshold provides a necessary but not sufficient condition and, for this reason, we will thus impose stronger detectability and resolvability conditions  (c.f.\ (\ref{eq:detect})-(\ref{eq:resolve})). 

Indeed, the mixing term in Eq.\ (\ref{eq:SNRmix}) plays an important role in degrading the naive expected precision on the QQNM parameters. For instance, in Fig.\ \ref{fig:ET} we obtain that $\rho_{440}=6.1$ and $\rho_{44Q}=19$ in ET, while the total SNR in the (44) angular harmonic is $\rho_{44}=16$,
according to Eq.\ (\ref{eq:SNRmix}). Due to the mixing between the (440) and $(44Q)$ QNMs, we find the SNR not to be high enough to be able to resolve independently the $(440)$ and $(44Q)$ modes, nor to detect non-vanishing QNM amplitudes with over $68\%$ confidence level for both modes. More generally, we will find that we can account for this mixing by imposing a higher threshold in the total (44) SNR of about $\rho_{44}\gtrsim 30$.

\subsection{Fisher formalism}
Next, we summarize the Fisher matrix formalism used to assess the precision at which QNM frequencies and amplitudes will be measured, and hence the confidence in the detectability of linear and quadratic QNMs.

Let us assume the ringdown model to be a function of a set of $N$ parameters $\{x^{i}\}$ with $i=1...N$. Typically, each QNM contributes with four parameters, given by amplitude magnitude $|A|$, amplitude phase $\phi$ such that $A=|A|\exp\{i\phi\}$, as well as real and imaginary frequencies $\omega_{R}$ and $\omega_I$ (for simplicity, we are omitting the labels $(\ell m n)$). For high enough SNR, the probability distribution of each parameter will be a Gaussian with a central value close to the real one, if no bias is present \cite{Vallisneri:2007ev} (as assumed here). 

The covariance matrix is a $N\times N$ matrix for the model parameters, which can be obtained from the Fisher matrix:
\begin{equation}\label{Fish_matrix}
    \Gamma_{ij}=\left\langle \frac{\partial h}{\partial x^i} \bigg|  \frac{\partial h}{\partial x^j} \right\rangle,
\end{equation}
which is in turn calculated as the frequency-domain overlap in Eq.\ (\ref{Overlap_eqn}) of the derivatives of the strain, evaluated at the true parameter values. The Fisher matrix relates to the expected Covariance matrix, $\mathbf{C}$, by:
\begin{eqnarray}
    \mathbf{C}= \Gamma^{-1}.
\end{eqnarray}
Then, we can read the expected root-mean-squared error on a given parameter $x^i$ as 
\begin{eqnarray}\label{sigma}
    \sigma_i^2=C_{ii},
\end{eqnarray}
while the correlation coefficient between two parameters $x^i$ and $x^j$ will be given by:
\begin{eqnarray}
    c_{ij}=\frac{C_{ij}}{\sqrt{C_{ii}C_{jj}}},
\end{eqnarray}
which ranges from 0 (uncorrelated, independent) to $\pm 1$ (perfectly correlated/anti-correlated).

Note that we could also estimate the error on the parameter $x^i$ as $\sigma_i^2=1/\Gamma_{ii}$, which would return the unmarginalized uncertainty (i.e.\ not accounting for parameter correlations). This difference will be important for the QQNM precision parameter estimation, as this mode will be indeed correlated with the (440) mode, and such correlation will greatly degrade the precision of some of the QQNM parameters.

In order to claim detectability of any given QNM with peak amplitude $A$, we will use the following detectability criterion:
\begin{equation}
    \sigma_{|A|} < |A|, \label{eq:detect}
\end{equation}
where $\sigma_{|A|}$ is the $68\%$ uncertainty on the amplitude, obtained with the Fisher matrix approach. 
This implies that a vanishing amplitude value (that is, the absence of that QNM) is outside the $68\%$ confidence region.

Furthermore, since we will be analyzing a ringdown model with multiple QNMs at the same time, we will use the Rayleigh criterion to evaluate their resolvability, i.e.\ whether different QNMs would be distinguishable from each other (as used in previous works e.g.\ \cite{Berti:2007zu, Bhagwat:2019dtm, JimenezForteza:2020cve}). For any pair of QNM frequencies, with real and imaginary differences $\Delta \omega_R$ and $\Delta \omega_R$, they will be claimed to be distinguishable if:
\begin{align}
|\Delta \omega_R| &> \textrm{max}(\sigma_{\omega_R}), \nonumber \\
|\Delta {\omega_I}| & > \textrm{max}(\sigma_{\omega_R}),\label{eq:resolve}
\end{align}
where $\sigma_{\omega_R}$ and $\sigma_{\omega_I}$ are the $68\%$ uncertainties in their measurements. 

In the linear ringdown theory, due to the orthogonality of the angular spherical  harmonics, $\Gamma$ and $\mathbf{C}$ can always be expressed in terms of block matrices, $\Gamma^{(\ell |m|)}$, one for each angular harmonic present in the ringdown model\footnote{Note this holds even if overtones are present. For each independent overtone $(\ell, m, n)$ included in the model, the dimensions of the corresponding $\Gamma^{(\ell |m|)}$ matrix would increase by four.}. In such a case, we would expect to have  three relevant blocks: a 4x4 matrix $\Gamma^{22}$ for the $(\ell |m|)=(22)$ harmonic with the (220) mode, a 4x4 matrix $\Gamma^{33}$ for the $(\ell |m|)=(33)$ harmonic with the (330) mode, and a 4x4 matrix $\Gamma^{44}$ for the $(\ell |m|)=(44)$ harmonic with the (440) mode. In this paper, when including the specific QQNM introduced above, additional mixing can occurs. We will specifically apply the Fisher matrix formalism to two scenarios:
\begin{itemize}
\item[(a)] Independent QQNM\\
We assume there are four independent QNMs in the model, each one contributing four parameters to the Fisher matrix. In other words, we consider the QQNM to be an independent QNM, which effectively contributes in analogous way as an overtone $(44n)$. 

As a result, $\Gamma^{44}$ will have dimensions 8x8, for the four parameters of the $(440)$ and $(44Q)$ modes. In Appendix \ref{app:fisher_indep} we provide the explicit expressions for all the elements of the Fisher matrix. 

\item[(b)] Dependent QQNM \\
We assume there are four QNMs in the model, with all the parameters of the $(44Q)$ mode fixed by their relationships to the $(220)$ mode, according to Eqs.\ (\ref{Qamp})-(\ref{Qfreq}).  As a consequence, there will only be three independent QNMs in the model. The QQNM is present, but does not add new parameters to those associated with the three linear QNMs.

As a result, the full Fisher matrix can be expressed in terms of a 4x4 $\Gamma^{44}$ matrix for the four parameters of the $(440)$ mode only, in addition to a 4x4 mixing matrix $\Gamma^{22x44}$ for the mixing between the $(220)$ and $(440)$ QNM parameters induced by the QQNM.
In Appendix \ref{app:fisher_dep} we provide the explicit expressions for all the elements of the Fisher components in this scenario. 
\end{itemize}

We note that Scenario (a) is the one considered in \cite{Yi:2024elj}, whereas Scenario (b) has not been analyzed before. Both cases describe different, complementary analyses. 
On the one hand, Scenario (a) describes a case in which the detection of the $(44Q)$ mode can be used to perform nonlinear spectroscopic tests of GR, by measuring the (220) and (44Q) parameters independently and comparing to the expected predictions in Eqs.\ (\ref{Qamp})-(\ref{Qfreq}).
On the other hand, Scenario (b) describes a case in which GR is assumed, and hence the $(44Q)$ mode is included to help improve precision on the (220) and (440) parameters. This will be useful to perform more accurate consistency tests of GR with the linear modes. Nonetheless, if the QQNM and linear QNM relationship were known in a given modified gravity theory, one could also use Scenario (b) to perform ringdown consistency tests on such a modified gravity theory.
It is also worth pointing out that, of the expressions relating the parameters of $(44Q)$ to those of $(220)$, Eq.\ (\ref{Qfreq}) on frequency is more robust than Eq.\ (\ref{Qamp}) on amplitude, in the sense that Eq.\ (\ref{Qfreq}) always holds, while the exact value in Eq.\ (\ref{Qamp}) varies with spin \cite{Khera:2024yrk}, and was obtained assuming the linear QNMs as the only source to the QQNM. However, it is known that general linear ringdown solutions can have non-trivial time evolutions in addition to the linear QNMs and those contributions may also affect the amplitude of the QQNM. Thus, one could contemplate fixing the frequency relation while relaxing the amplitude one (for instance, by giving the latter a spread). We expect the resulting conclusions regarding signal-to-noise would be intermediate between Scenarios (a) and (b).

\section{Results: Independent QQNM}\label{sec:results1}
In this section, we show the results of the Fisher forecast in scenario (a), when assuming that the QQNM parameters are independent from those of the linear QNMs.

\subsection{Stellar-mass BBHs}\label{sec:indep-SMBH}

We start by analyzing an event similar to GW150914, by using the SXS simulation BBH:0305 with $z=0.093$, with final remnant mass $M=61M_\odot$ and spin $\chi=0.69$. Considering the modes (220), (330), (440) and (44Q), this event has a total ringdown SNR of $\rho= 200$ in ET and $\rho=139$ in CE, and the values for each QNM in the model are given in Table \ref{Table_SNR_1}. 
\begin{table}[h!]
\centering %\vspace{-1em}
\begin{tabular}{|c|cc|} 
\hline
Single mode $\rho$ & ET ($z=0.093$)  &  CE ($z=0.093$) \\\hline
$\rho_{220}$ & $199$ &  $138.26$ \\\hline
$\rho_{330}$ & $11.3$ &  $7.83$ \\\hline
$\rho_{440}$ & $5.68$ &  $3.97$ \\\hline
$\rho_{44Q}$ & $18.1$ &  $12.6$ \\\hline
$\rho_{44}$ & $15.2$ &  $10.6$ \\\hline
\end{tabular}
\caption{Single mode SNRs for a GW150914-like event (using simulation SXS:BBH:0305). The $\rho_{44}$ value is the total SNR in the $(\ell,|m|)=(4, 4)$ angular harmonics, according to Eq.\ (\ref{eq:SNRmix}). }\label{Table_SNR_1}
\end{table}
Most of this SNR comes from the dominant (220) mode. From Fig.\ \ref{fig:ET} we can see that, for this binary, the QNM frequencies lie in the higher-frequency range of the sensitivity spectrum, where ET has higher sensitivity than CE, which explains the differences in SNR obtained here. 

As one would have expected, based on the decay rates and amplitudes, the (220) mode has higher SNR than the rest. Note that the difference in SNR between the (220) and $(44Q)$ QNMs is expected since the QQNM has nearly one order of magnitude peak amplitude smaller than the (220) mode (c.f.\ Table \ref{table:QNM_fits}), and it decays twice as fast.

Crucially, for all the cases studied in this section, we find the SNR of the QQNM to be always larger than the (440) and (330) modes. This happens because, even though the (330) and (440) modes decay slower than the $(44Q)$ mode, the QQNM has a larger peak amplitude (c.f.\ Table \ref{table:QNM_fits}) that makes it very  relevant in the total signal, and in future spectroscopic test of gravity.

For this GW150914-like event, we  find that neither the detectability nor the resolvability criterion in Eqs.\ (\ref{eq:detect})-(\ref{eq:resolve}) is satisfied: the (440) and (44Q) modes have overlapping error bars and amplitudes compatible with vanishing values. This tells us that the threshold $\rho_{44Q}>8$ is simply a necessary condition. For this reason, we reduce the distance of this event, such that both detectability and resolvability criteria are simultaneously satisfied. Generally, we find Eq.\ (\ref{eq:detect}) to be more restrictive than (\ref{eq:resolve}), and we found the maximum distances that satisfy both conditions to be $z=0.045$ for ET and $z=0.031$ for CE. These maximum redshifts are driven by the precision on the (440) QNM amplitude, which is the worst measured parameter of all. This means that the same maximum redshifts would be obtained in an analogous analysis of a purely linear ringdown model (i.e.\ without the QQNM in the model).

We find the total SNR is $400$ for ET (using $z=0.045$) and $396$ for CE (using $z=0.031$), and the individual mode SNRs are shown in Table \ref{Table_SNR_1v2}. From this Table we see that for all modes to be resolved and detectable, we need the lowest individual QNM SNR to be at least 10 (in this binary event, this is the case for the (440) mode), and a total $(\ell,|m|)=(4,4)$ SNR of at least 30.

\begin{table}[h!]
\centering %\vspace{-1em}
\begin{tabular}{|c|cc|} 
\hline
Single mode $\rho$ & ET ($z=0.045$)  &  CE ($z=0.031$) \\\hline
$\rho_{220}$ & $399$ &  $396$ \\\hline
$\rho_{330}$ & $22.2$ &  $22.2$ \\\hline
$\rho_{440}$ & $11.1$ &  $11.1$ \\\hline
$\rho_{44Q}$ & $35.5$ &  $35.3$ \\\hline
$\rho_{44}$ & $29.8$ &  $29.7$ \\\hline
\end{tabular}
\caption{
Single mode SNRs for a GW150914-like event (using simulation SXS:BBH:0305), as in  Table \ref{Table_SNR_1} but for smaller redshift so that the event satisfies the detectability and resolvability criteria in each detector.} \label{Table_SNR_1v2}
\end{table}

We emphasize that since the $(440)$ mode has lower SNR than the $(44Q)$ mode, the maximum redshifts quoted here to resolve the detectability and resolvability criteria could be relaxed considerably if only the $(44Q)$ is desired to be measured confidently. In such a case, the resolvability and detectability criteria would be satisfied for the $(220)$, $(330)$, and $(44Q)$ modes only.
If we allowed imprecise $(440)$ measurements, the redshifts could be extended to $z = 0.23$ for ET and $z = 0.15$  for CE, encompassing then the event GW150914. For these limiting redshifts, the QQNM is required to have an SNR of at least 6 to be detectable.

\begin{figure}[h!]
\centering
\includegraphics[width=0.99\linewidth]{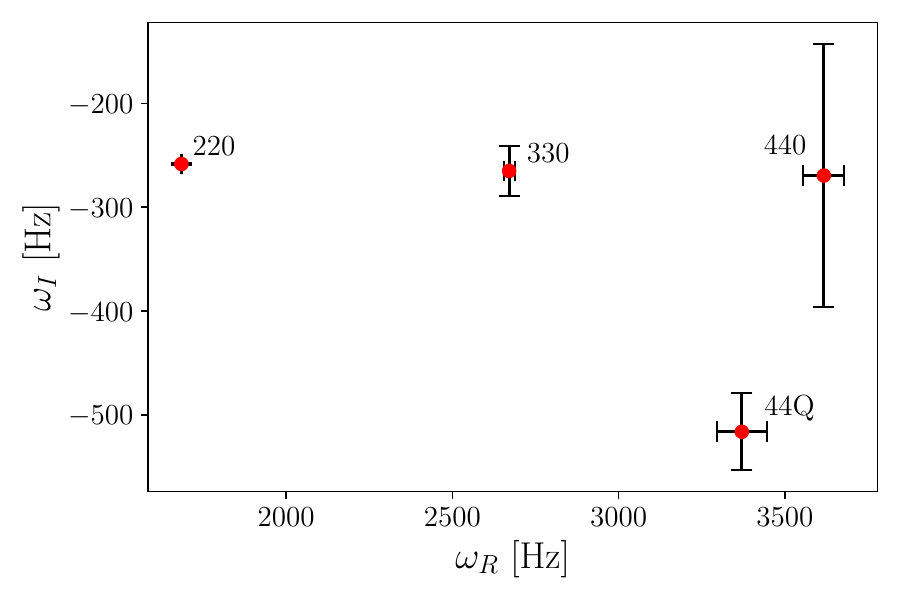}
\caption{\label{fig:0305Lin} Real and imaginary components of the QNM frequencies (in Hz) and their $1\sigma$ uncertainty, for a GW150914-like event at $z=0.045$ in ET.}
\end{figure}

Fig.\ \ref{fig:0305Lin}  shows the results for the four QNM frequencies and their $1\sigma$ uncertainty (in Hertz) in this mock event at $z=0.045$ in ET. From here we see that all of these modes satisfy the criteria. Furthermore, according to the SNR results, we find the uncertainties to be smallest for the (220) mode, followed by the (330) mode, then the ($44Q$) mode, with the worst ones for the (440) mode.

Quantitatively, we obtain the following fractional marginalized uncertainties (using Eq.\ (\ref{sigma})) in the QNM frequencies, both its real and imaginary parts, using ET at $z = 0.045$,
\begin{align}
&\frac{\sigma_{\omega_{R,220}}}{ \omega_{R,220}}=0.055\% , \quad
    \frac{\sigma_{\omega_{I, 220}}}{ \omega_{I,220}}=0.53\% ; \label{ET_Dw220}\\
   & \frac{\sigma_{\omega_{R, 330}}}{ \omega_{R, 330}}=0.63\% , \quad
    \frac{\sigma_{\omega_{I, 330}}}{ \omega_{I, 330}}=9.0\% ; \label{ET_Dw330}\\
&\frac{\sigma_{\omega_{R, 440}}}{ \omega_{R, 440}}=1.7\% , \quad
\frac{\sigma_{\omega_{I, 440}}}{ \omega_{I, 440}}=47\% ;\label{ET_Dw440}\\
     &\frac{\sigma_{\omega_{R, 44Q}}}{ \omega_{R, 44Q}}=2.3\% ,\quad
     \frac{\sigma_{\omega_{I, 44Q}}}{ \omega_{I, 44Q}}=7.1\%  . \label{ET_Dw44Q}
\end{align}
From these results, we see that the real part of each mode is constrained about 10 times better than their corresponding imaginary parts, in agreement with the previous QQNM study in \cite{Yi:2024elj} as well as linear QNM studies (e.g.\ in \cite{Branchesi:2023mws}. We emphasize that the QQNM is constrained to a comparable or better precision to the linear higher angular harmonics. For the mock event with $z=0.031$ in CE, we find similar results to those in Eqs.\ (\ref{ET_Dw220})-(\ref{ET_Dw44Q}), differing in a factor less than unity. 

\begin{figure}[h!]
\centering
\includegraphics[width=0.95\linewidth]{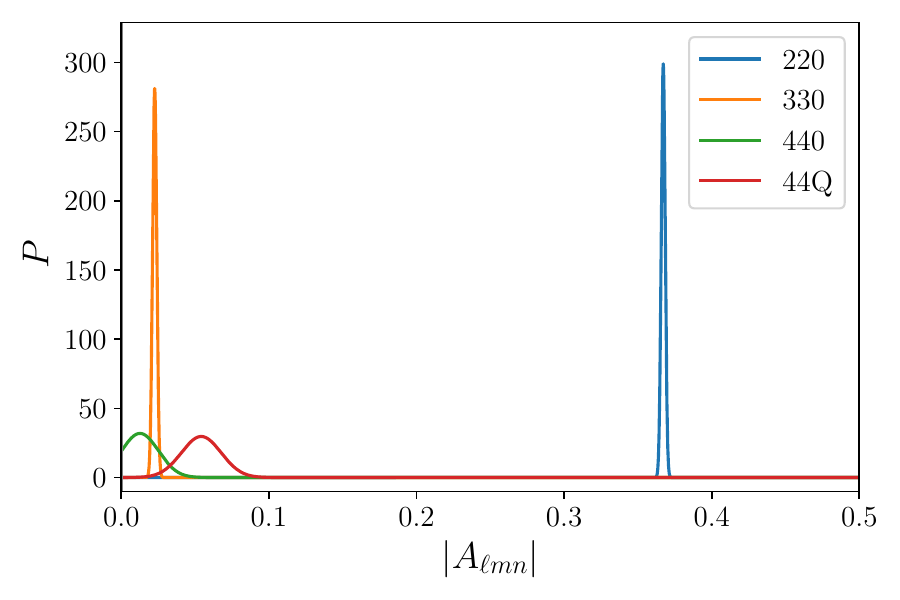}
\caption{\label{fig:0305Alin} Normalized probability distribution function of the peak amplitude magnitudes $|A|$ for a GW150914-like event at $z=0.045$ in ET.}
\end{figure}

Fig.\ \ref{fig:0305Alin} shows the results for the four QNM peak amplitudes, and their expected probability distributions in this mock event at $z=0.045$ in ET. While the (220), (330) and $(44Q)$ clearly satisfy the detectability criteria, the mode (440) is a limiting case. Therefore, this choice of $z=0.045$ is the maximum distance at which we would confidently claim detection of the four modes. Note that this maximum redshift is expected to vary with mass \cite{Ota:2021ypb}, so the values presented here are used as reference for stellar-mass binaries.

For the amplitudes, we obtain the following quantitative results:
\begin{align}
     &\frac{\sigma_{|A_{220}|}}{|A_{220}|}=0.36\% ,\quad  
   \frac{\sigma_{|A_{330}|}}{|A_{330}|}=6.3\% ; \label{ET_DA23}\\
      &\frac{\sigma_{|A_{440}|}}{|A_{440}|}=37\% , \quad  \frac{\sigma_{|A_{44Q}|}}{|A_{44Q}|}=30\%  .\label{ET_DA44}
\end{align}
As expected, for nearly-equal mass stellar-mass binaries, the amplitude of the (220) angular harmonic can be measured to percent-level precision, whereas those of the higher harmonics are about ten times less precise. Importantly, the amplitude of the $(44Q)$ QQNM is measured to slightly better precision than that of the $(440)$ linear mode. These estimates of the amplitude precision are also in agreement with past works \cite{Yi:2024elj,Branchesi:2023mws}.

From Tables \ref{Table_SNR_1} and \ref{Table_SNR_1v2} we see that the QQNM has considerably higher SNR than the (330) and (440) modes \footnote{If one were interested in unequal mass mergers, based on fittings to numerical simulations \cite{Mitman:2022qdl}, we expect this result to revert and make the (330) and (440) modes to have higher SNR than the QQNM mode.}. Nonetheless, from Eqs.\ (\ref{ET_Dw220})-(\ref{ET_DA44}) we see that this SNR relationship does not always directly predict which QNM parameters will be better. On the one hand,  when comparing the (440) mode to the QQNM, we have found that the amplitude and imaginary frequency of the QQNM mode are measured better, while its real frequency is measured worst. On the other hand, when comparing the (330) mode to the QQNM, we have found that the amplitude and real frequency of the (330) mode are measured better, and only its imaginary frequency is measured worst. 

These results are due to the fact that the (440) and ($44Q$) parameters are degenerate, which degrades the naive parameter precision expected for them. Indeed, if we calculate the unmarginalized errors on the QNM parameters (simply as $\sigma_i^2=1/\Gamma_{ii}$) we find most of the parameters of the QQNM and the (440) mode to be measured at a comparable or better precision than those of the (330) mode. For instance, for the amplitudes we obtain $\sigma_{|A_{44Q}|}/|A_{44Q}|=3.4\%$ and $\sigma_{|A_{440}|}/|A_{440}|=3.3\%$, compared to $\sigma_{|A_{330}|}/|A_{330}|=4.50\%$ for the unmarginalized errors. These errors get degraded by nearly a factor of 10 for the (440) and $(44Q)$ QNMs when accounting for their mixing.

Finally, we conclude this subsection highlighting that the results obtained here are expected to be applicable for similar binaries (quasi-circular, nearly equal-mass, small  spins) with different redshifts and total mass. In order to do that, we quote the expected fractional precision, normalized by the corresponding mode SNR as follows:
\begin{align}
     &\frac{\sigma_{\omega_{R,220}}}{ \omega_{R,220}}=0.22\% \left(\frac{100}{\rho_{22}}\right); \quad \frac{\sigma_{\omega_{I, 220}}}{ \omega_{I,220}}=2.1\% \left(\frac{100}{\rho_{22}}\right), \label{Dw_220_SNR}\\
     &\frac{\sigma_{\omega_{R, 330}}}{ \omega_{R, 330}}=1.4\% \left(\frac{10}{\rho_{33}}\right); \quad \frac{\sigma_{\omega_{I, 330}}}{ \omega_{I, 330}}=20\% \left(\frac{10}{\rho_{33}}\right), \\
      &\frac{\sigma_{\omega_{R, 440}}}{ \omega_{R, 440}}=1.9\% \left(\frac{10}{\rho_{44}}\right); \quad \frac{\sigma_{\omega_{I, 440}}}{ \omega_{I, 440}}=52\% \left(\frac{10}{\rho_{44}}\right),\\
      & \frac{\sigma_{\omega_{R, 44Q}}}{ \omega_{R, 44Q}}=8.0\% \left(\frac{10}{\rho_{44}}\right), \quad \frac{\sigma_{\omega_{I, 44Q}}}{ \omega_{I, 44Q}}=25\% \left(\frac{10}{\rho_{44}}\right).
 \end{align}
 and 
\begin{align}
&  \frac{\sigma_{|A_{220}|}}{|A_{220}|}=1.4\% \left(\frac{100}{\rho_{22}}\right)   ; \quad  \frac{\sigma_{|A_{330}|}}{|A_{330}|}=14\% \left(\frac{10}{\rho_{33}}\right) , \\
&  \frac{\sigma_{|A_{440}|}}{|A_{440}|}=110\% \left(\frac{10}{\rho_{44}}\right) ; \quad  \frac{\sigma_{|A_{44Q}|}}{|A_{44Q}|}=90\% \left(\frac{10}{\rho_{44}}\right) \label{Dw_44_SNR}.
\end{align}
This way of normalizing the errors is different from the one typically used in the literature, where errors are normalized by total SNR instead of mode SNRs. However, we find this normalization to be more informative as it will be valid for binaries that not only change the overall SNR but also change the relative mode SNR. For the linear modes, these results are consistent with those quoted in \cite{Branchesi:2023mws}. These results are thus expected to be rather generic. Even if the ringdown model was extended to include overtones of the (22) angular harmonic, none of the results for the (330), (440), and QQNM would be affected since they are uncorrelated with the (22) angular harmonic.

\subsection{Intermediate-mass BBHs}

In this section, we consider the merger of heavier, spinning BHs, akin to GW190521 \cite{LIGOScientific:2020iuh}. Such an event was consistent with a redshift $z=0.82$, individual masses $m_1=85M_\odot$ and $m_2=66M_\odot$, and thus a mass ratio of $q=1.29$, with a remnant mass $M=142M_\odot$. 
Importantly, the data was consistent with two aligned dimensionless spins  $\chi_1=0.69$, $\chi_2=0.73$, suggesting such BHs to be of second generation. While these types of binaries are less common than first-generation BHs with a typical merger rate of about $0.01-10\text{Gpc}^{-3}\text{yr}^{-1}$ \cite{LIGOScientific:2021tfm,KAGRA:2021duu}, future detectors CE and ET are expected to observe high redshift mergers, up to $z\sim 10$, potentially observing up to a few thousand intermediate-mass merger per year \cite{Gair:2010dx, Rasskazov:2019tgb, Fragione:2022avp}. 

In this section, we will use the simulation SXS:BBH:0507 to represent an intermediate-mass second-generation BH binary (IMBBH) system. We will initially choose $z=0.82$ and total mass $M=142M_\odot$, consistent with the event GW190521. While this event showed some mild evidence for precession, our mock event will not account for precession. The inclusion of precession would change the dominant set of angular harmonics during the ringdown, as discussed in \cite{Siegel:2023lxl}. Another difference will be the fact this simulation allows for one large spin instead of two. We will choose $\chi_1=0.8$ and $\chi_2=0.4$ as representative values.

This mock event will yield different results as the one analyzed in Sec.\ \ref{sec:indep-SMBH} since, as discussed in \cite{Cheung:2023vki, Yi:2024elj}, binaries with higher remnant spins (and thus individual spins) lead to larger QQNMs amplitudes, and here we confirm these results. While BBH:0305 predicts a relative QQNM amplitude of $|A_{44Q}/A_{220}|=0.098$ at $10M$ after the merger, BBH:0507 predicts $|A_{44Q}/A_{220}|=0.12$ (see Table \ref{table:QNM_fits}). 

For this mock event, with $z=0.82$ and $M=142M_\odot$, we obtain that the total SNR in the ringdown, starting from $10M$ after the merger, is $\rho=113$ for ET and $\rho=104$ for CE. More specifically, in such a case we obtain $\rho_{44}=26.5$ for ET and $\rho_{44}=20.0$ for CE, and we find this SNR to be too low and break both resolvability and detectability criteria. Similarly to the stellar-mass binary mock event in Sec.\ \ref{sec:indep-SMBH}, here the $(440)$ mode is the worst measured one and the responsible for the breaking of both criteria. 
For this reason, we again lower the redshift to $z = 0.51$ for ET, and $z= 0.32$ for CE such that both criteria are satisfied. In such limiting cases, the total SNR is found to be $\rho= 164$ for ET, and $\rho= 201$ for CE, while the individual mode SNRs are quoted in Table \ref{Table_SNR_2}. 

\begin{table}[h!]
\centering %\vspace{-1em}
\begin{tabular}{|c|cc|} 
\hline
Single mode $\rho$ & ET ($z=0.51$)  &  CE ($z=0.32$) \\\hline
$\rho_{220}$ & $159$ &  $197$ \\\hline
$\rho_{330}$ & $10.7$ &  $11.6$ \\\hline
$\rho_{440}$ & $13.5$ &  $13.5$ \\\hline
$\rho_{44Q}$ & $45.7$ &  $46.6$ \\\hline
$\rho_{44}$ & $37.0$ &  $37.8$ \\\hline
\end{tabular}
\caption{Single mode SNRs for an intermediate-mass binary such as GW190521 (using simulation SXS:BBH:0507). The values of the redshift are the largest for which the event satisfies the detectability and resolvability criteria in each detector.}\label{Table_SNR_2}
\end{table}

In Table \ref{Table_SNR_2} we again find the QQNM to have considerably higher SNR than the (330) and (440) linear QNMs, in this case by a factor of four. In addition, we now find the lowest mode SNR to be over 13, and the total (4,4) SNR to be over 35 in order to satisfy both detectability and resolvability criteria. These SNR limiting values are comparable to those of the BBH:0305 mock event studied in the previous section, yet the limiting redshift is about 10 times higher than the one found in the previous section due to the higher total mass of this mock event. Indeed, the maximum redshift is expected to vary with total mass \cite{Ota:2021ypb}.

\begin{figure}[h!]
\centering
\includegraphics[width=0.95\linewidth]{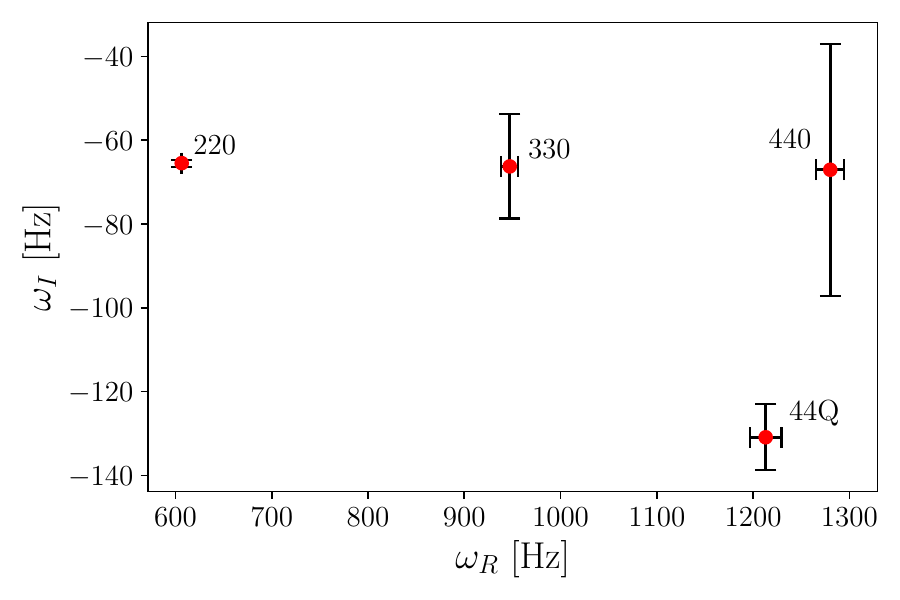}
\caption{\label{fig:0507Lin} Real and imaginary components of the QNM frequencies (in Hz) and their $1\sigma$ uncertainty for a GW190521-like event in ET at $z=0.51$. }
\end{figure}

\begin{figure}[h!]
\centering
\includegraphics[width=0.95\linewidth]{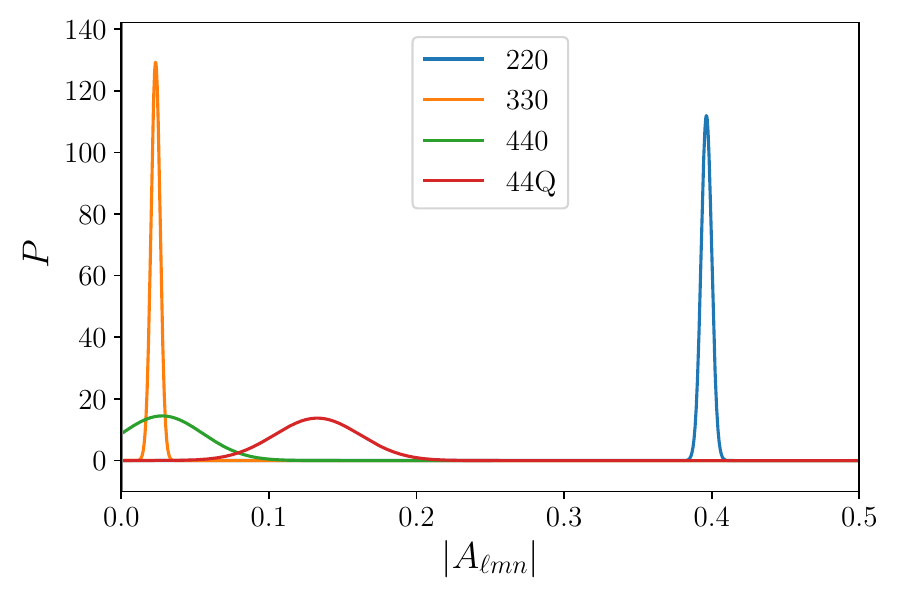}
\caption{\label{fig:0507Alin} Normalized probability distribution function of amplitudes $|A|$, for a GW190521-like event for ET at $z=0.51$. }
\end{figure}

Similarly to Sec.\ \ref{sec:indep-SMBH}, the results on the precision of frequency and amplitudes for this IMBBH system are shown in Figs.\ \ref{fig:0507Lin} and \ref{fig:0507Alin}, for the ET case (with $z= 0.51$). We again find the resolvability criteria to be well satisfied, while the detectability criteria for the $(440)$ mode is barely satisfied. 

Quantitatively, we find the following fractional uncertainties  in the QNM frequencies, for this event with $z=0.51$ in ET,
\begin{align}
    &\frac{\sigma_{\omega_{R,220}}}{ \omega_{R,220}}=0.094\% ; \quad
    \frac{\sigma_{\omega_{I, 220}}}{ \omega_{I,220}}=1.3\% ; \\
    &\frac{\sigma_{\omega_{R, 330}}}{ \omega_{R, 330}}=0.94\% ; \quad
    \frac{\sigma_{\omega_{I, 330}}}{ \omega_{I, 330}}=19\% ; \label{Dw_IMBH330}\\
     &\frac{\sigma_{\omega_{R, 440}}}{ \omega_{R, 440}}=1.1\%   ; \quad
     \frac{\sigma_{\omega_{I, 440}}}{ \omega_{I, 440}}=45\%  ;\\
     &\frac{\sigma_{\omega_{R, 44Q}}}{ \omega_{R, 44Q}}=1.4\%  ; \quad
     \frac{\sigma_{\omega_{I, 44Q}}}{ \omega_{I, 44Q}}=6.0\% .
\end{align}
We also confirm that for CE the fractional uncertainties are similar to the ones for ET, with differences less than a factor of unity. 
We now see that, even though the QQNM has four times higher SNR than (330) and (440), the precision on the real frequency is very similar for the three modes, while the precision for the imaginary frequency is considerably better for the QQNM. 

For the amplitudes, we obtain the following quantitative results, for ET,
\begin{align}
   \frac{\sigma_{|A_{220}|}}{|A_{220}|}=0.90\%  ; \quad
   \frac{\sigma_{|A_{330}|}}{|A_{330}|}=13\%  ; \label{ET_DA23_2}\\
    \frac{\sigma_{|A_{440}|}}{|A_{440}|}=99\% ; \quad  \frac{\sigma_{|A_{44Q}|}}{|A_{44Q}|}=22\% .\label{ET_DA44_2}
\end{align}
where we find the amplitude of the (330) mode to be measured better than that of the QQNM, while the latter has almost five times better precision than the (440) mode. 

In Fig.\ \ref{fig:0507Alin} we see that the detectability condition is satisfied for the (220), (330) and $(44Q)$ modes, and barely satisfied for the (440) mode. From this plot, we also visually confirm that the relative amplitude of the QQNM is larger than for the non-spinning BBHs in Fig.\ \ref{fig:0305Alin}, which makes the QQNM more easily to detect (i.e.\ distinguishable from zero) and thus potentially more useful for future spectroscopic tests of gravity.

Finally, similar to the precious section, we quote the fractional precision normalized by mode SNR, in order to be applicable to binaries with different masses and redshifts:
\begin{align}
    &\frac{\sigma_{\omega_{R,220}}}{ \omega_{R,220}}=0.15\% \left(\frac{100}{\rho_{22}}\right); \quad \frac{\sigma_{\omega_{I, 220}}}{ \omega_{I,220}}=2.1\% \left(\frac{100}{\rho_{22}}\right) , \\
    &\frac{\sigma_{\omega_{R, 330}}}{ \omega_{R, 330}}=1.0\% \left(\frac{10}{\rho_{33}}\right) ; \quad \frac{\sigma_{\omega_{I, 330}}}{ \omega_{I, 330}}=20\% \left(\frac{10}{\rho_{33}}\right) , \\
     &\frac{\sigma_{\omega_{R, 440}}}{ \omega_{R, 440}}=4.1\% \left(\frac{10}{\rho_{44}}\right) ; \quad \frac{\sigma_{\omega_{I, 440}}}{ \omega_{I, 440}}=167\% \left(\frac{10}{\rho_{44}}\right) ,\\
     & \frac{\sigma_{\omega_{R, 44Q}}}{ \omega_{R, 44Q}}=5.2\% \left(\frac{10}{\rho_{44}}\right) , \quad \frac{\sigma_{\omega_{I, 44Q}}}{ \omega_{I, 44Q}}=22\% \left(\frac{10}{\rho_{44}}\right) ,
\end{align}
and
\begin{align}
   &  \frac{\sigma_{|A_{220}|}}{|A_{220}|}=1.4\% \left(\frac{100}{\rho_{22}}\right)  ; \quad  \frac{\sigma_{|A_{330}|}}{|A_{330}|}=14\% \left(\frac{10}{\rho_{33}}\right) , \\
    &  \frac{\sigma_{|A_{440}|}}{|A_{440}|}=368\% \left(\frac{10}{\rho_{44}}\right) ; \quad  \frac{\sigma_{|A_{44Q}|}}{|A_{44Q}|}=82\% \left(\frac{10}{\rho_{44}}\right) .
\end{align}
Since we are normalizing by mode SNR, as opposed to total SNR, these results for the (220) and (330) modes are comparable (with differences less than a factor of unity) to the ones found for the representative first generation BBH event in Eqs.\ (\ref{Dw_220_SNR})-(\ref{Dw_44_SNR}), confirming the robustness of these estimations. Nonetheless, some of the results for the (440) and $(44Q)$ modes change by more than a factor of unity, because their relative contribution to the signal is different from the one for the first generation BBHs, yet we are still normalizing by the total $\rho_{44}$.  

\section{Results: Dependent QQNM}\label{sec:results2}

In this section, we show how the previous results change if we were to use the GR predictions that relate the QQNM to its parent linear QNMs, in Eqs.\ (\ref{Qamp})-(\ref{Qfreq}). In this case, the $(44Q)$ QNM parameters will all be dependent on the (220) ones thus improving the measurements of the (220) and (440) linear modes. Therefore, we will not use the fitted value $A_{44Q}$ in Table \ref{table:QNM_fits} for the QQNM, and instead we will derive its value from Eq.\ (\ref{Qamp}) which yields $|A_{44Q}|=2.7\times 10^{-2}$ for BBH:0305, and $|A_{44Q}|=4.3 \times 10^{-2}$ for BBH:0507. Since Eq.\ (\ref{Qamp}) may vary with spin, albeit weakly, the results presented here should be considered as estimates. More accurate calculations could be performed once the constant factors in Eq.\ (\ref{Qamp}) are updated for non-zero spins.

Notice that all results for the (330) QNM are the same as the ones quoted in the previous section, since the QQNM studied in this paper is independent and uncorrelated with the QQNM. For this reason, we do not quote explicitly the results for the (330) mode in this section.

Additionally, all the SNR values quoted in the previous section will hold again here. The signal is assumed to be the same, the only difference will be the number of independent parameters used in the ringdown model to be fitted, and their expected uncertainties. In Appendix \ref{app:fisher_dep} we present the explicit expressions used in this section for the fisher forecast for the only independent QNM parameters: those of the (220), (330), and (440) linear QNMs.

\subsection{Stellar-mass BBHs}
\begin{figure}[h!]
\centering
\includegraphics[width=0.99\linewidth]{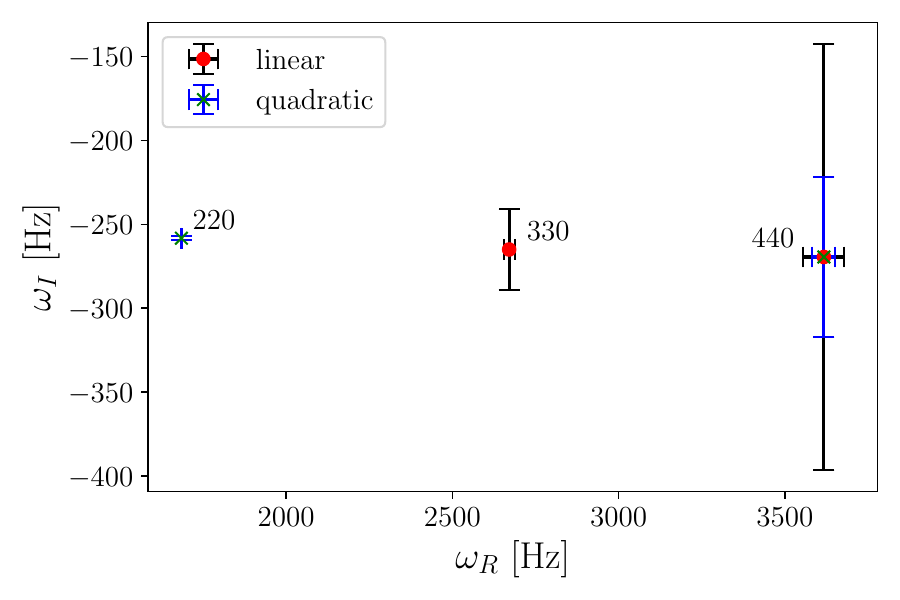}
\caption{\label{fig:0305QDep} Real and imaginary components of the QNM frequencies (in Hz) and their $1\sigma$ uncertainty, for a GW150914-like event for ET at $z=0.045$. The blue (black) error bars correspond to the dependent (independent) QQNM scenario. } 
\end{figure}

\begin{figure}[h!]
\centering
\includegraphics[width=0.99\linewidth]{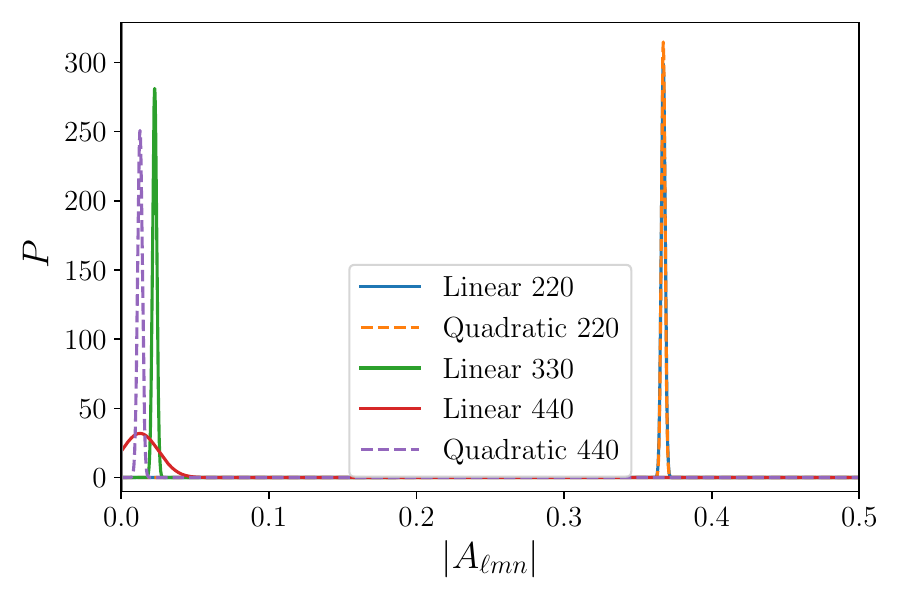}
\caption{\label{fig:0305ADep} Normalized probability distribution function for QNM amplitudes $|A|$, for a GW150914-like event for ET at $z=0.045$. Dashed (solid) lines correspond to the dependent (independent) QQNM scenario.} 
\end{figure}

For the same event of a stellar-mass binary described in Sec.\ \ref{sec:indep-SMBH} we show the results with a dependent QQNM in Figs.\ \ref{fig:0305QDep} and \ref{fig:0305ADep}. In Fig.\ \ref{fig:0305QDep} we compare the results of the uncertainties on the frequencies of the $(220)$, $(330)$, and $(440)$ QNM parameters. In black, we show the same error bars obtained in Sec.\ \ref{sec:indep-SMBH} (c.f.\ Fig.\ \ref{fig:0305Lin}) while in blue we show the new error bars assuming a dependent QQNM.

In Fig.\ \ref{fig:0305QDep} we see that the precision on the QNM frequency parameters, both real and imaginary parts, improve mostly for the (440) mode. This happens due to the fact that this mode is considerably less correlated with the rest of the ringdown model parameters in this setup. 
Additionally, we find that the precision on the (220) frequency parameters is only slightly improved, since the information gained by adding the $(44Q)$ mode is low given its suppressed relative amplitude and faster decay time, when compared to the (220) mode.

Quantitatively, we obtain the following fractional precision on the QNM frequencies, for the GW150914-like event with $z=0.045$ in ET,
\begin{align}
    &\frac{\sigma_{\omega_{R,220}}}{ \omega_{R,220}}=0.056\%; \quad \frac{\sigma_{\omega_{I, 220}}}{ \omega_{I,220}}=0.50\% , \\
     &\frac{\sigma_{\omega_{R, 440}}}{ \omega_{R, 440}}=0.94\% ; \quad \frac{\sigma_{\omega_{I, 440}}}{ \omega_{I, 440}}=18\% ,
\end{align}
With these results, we confirm that the improvement on the (220) mode frequency parameters is less than $0.1\%$ (compare to Eq.\ (\ref{ET_Dw220})) for the (220) mode, which will be irrelevant given that the expected observational errors are larger than this value. For the (440) mode, the improvement is a factor of roughly $2$ (compare to Eq.\ (\ref{ET_Dw440})). For the event considered here, 
%since the SNR of the (44) angular harmonic is $\rho_{44}=29.70$ and the (33) mode is $\rho_{33}=22.17$, 
we find that the parameters of the (440) QNM are now measured to a comparable precision to those of the (330) mode, with the latter being still slightly better than the former (compare to Eq.\ (\ref{ET_Dw330})).

In Fig.\ \ref{fig:0305ADep} we show the precision on the amplitude of the (220), (330), and (440) linear QNMs. In solid lines we show the result obtained in the previous section (see Fig.\ \ref{fig:0305Alin}), and in dashed lines we show the result assuming a dependent QQNM. Visually, we confirm similar results as those for the frequency: the greatest improvement is for the (440) mode.

Quantitatively, we find the following fractional uncertainties in the QNM amplitudes:
\begin{align}
   &  \frac{\sigma_{|A_{220}|}}{|A_{220}|}=0.35\%   ; \quad  \frac{\sigma_{|A_{440}|}}{|A_{440}|}= 13\% .
\end{align}
Comparing to the results of the previous section, we find an improvement in the precision less than $0.1\%$ for the (220) mode, and a factor of over 2 for the (440) mode. 

Since now the (440) parameters are measured considerably better, in this model the detectability and resolvability criteria are more easily satisfied. We find the maximum distance satisfying such criteria to be $z = 0.60$ for ET and $z= 0.33$ for CE, in this GW150914-like event with $M=61M_\odot$. This is one order of magnitude improvement over the maximum distance obtained in the previous section ($z=0.045$ for ET and $z=0.031$ for CE). Therefore, we expect many more BBH events for which these four QNMs could be observationally confirmed in the future, given that the detection rates can grow considerably between redshift 0 and 1 \cite{LIGOScientific:2020kqk, Fishbach:2021yvy}.
Indeed, the event GW150914 had a redshift of $z=0.093$, which would easily allow for a confident detection of the three linear modes with both ET and CE, and therefore a single such events could prove incredibly useful for consistency spectroscopic tests of gravity. 

As in the previous section, we provide the final fractional errors normalized to SNR, so that they can be used for estimating errors in similar binaries with different masses and redshifts: 

\begin{align}
    &\frac{\sigma_{\omega_{R,220}}}{ \omega_{R,220}}=0.22\% \left(\frac{100}{\rho_{22}}\right); \quad \frac{\sigma_{\omega_{I, 220}}}{ \omega_{I,220}}=2.0\%\left(\frac{100}{\rho_{22}}\right) , \\
     &\frac{\sigma_{\omega_{R, 440}}}{ \omega_{R, 440}}=2.8\% \left(\frac{10}{\rho_{44}}\right); \quad \frac{\sigma_{\omega_{I, 440}}}{ \omega_{I, 440}}=53\% \left(\frac{10}{\rho_{44}}\right),
\end{align}
and
\begin{align}
   &  \frac{\sigma_{|A_{220}|}}{|A_{220}|}=1.4\% \left(\frac{100}{\rho_{22}}\right)  ; \quad  \frac{\sigma_{|A_{440}|}}{|A_{440}|}= 37\% \left(\frac{10}{\rho_{44}}\right).
\end{align}
We again emphasize that because of the way we have normalized the results, we expect them to be rather generic. Furthermore, even if we were to extend the ringdown model to include overtones of the (22) angular harmonic, the results for the (330) and (440) modes are not expected to change considerably. This is because, while technically there will be a non-vanishing correlation between the (440) QNM and such overtones (indirectly via the QQNM), we expect the correlation to be negligible as we have found with the (220) mode.

\subsection{Intermediate-mas BBHs}
\begin{figure}[h!]
\centering
\includegraphics[width=0.99\linewidth]{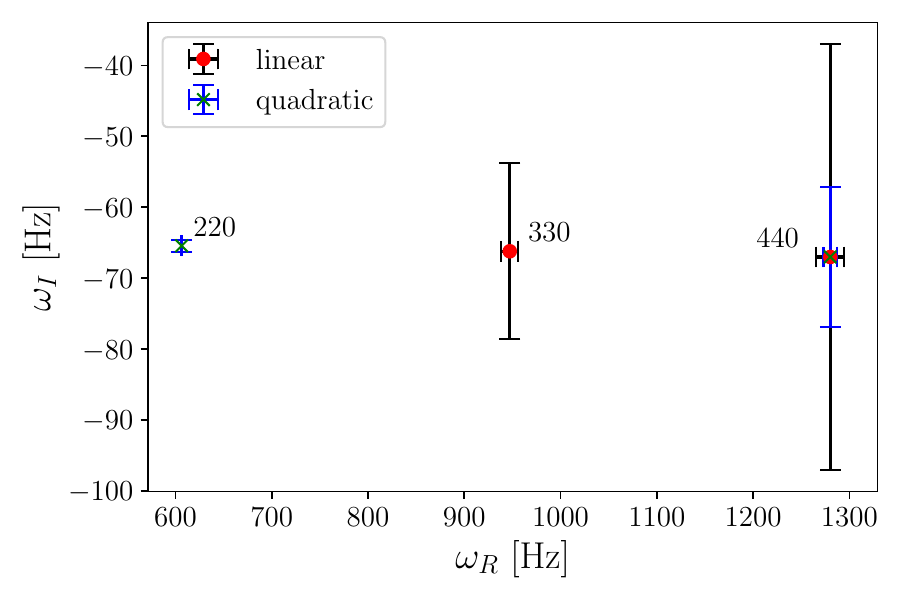}
\caption{\label{fig:0507QDep} Real and imaginary components of the QNM frequencies (in Hz) and their $1\sigma$ uncertainty, for a GW190521-like event for ET at $z=0.51$. The blue (black) error bars correspond to the dependent (independent) QQNM scenario.} 
\end{figure}

\begin{figure}[h!]
\centering
\includegraphics[width=0.99\linewidth]{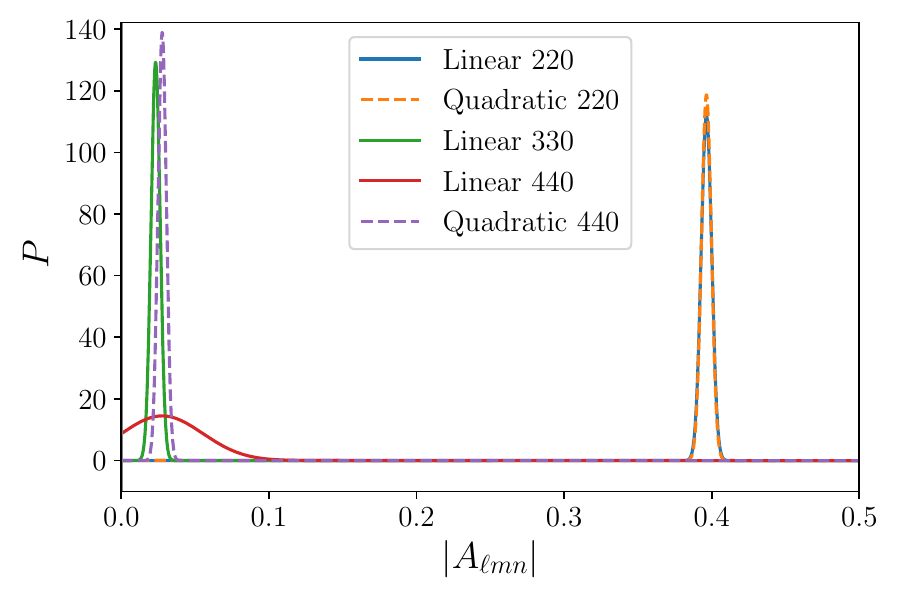}
\caption{\label{fig:0507ADep} Normalized probability distribution function for QNM amplitudes $|A|$, for a GW190521-like event for ET at $z=0.51$. Dashed (solid) lines correspond to the dependent (independent) QQNM scenario. } 
\end{figure}

Next, we show the results for our representative IMBBH event, similar to GW190521. In Figs.\ \ref{fig:0507QDep} and \ref{fig:0507ADep} we show the results for the frequency and amplitude QNM precision for an event at $z=0.51$ in ET, assuming a dependent QQNM. The results are qualitatively similar with those of the GW150914-like event, in the sense that the biggest improvement is seen in the (440) QNM parameters.

Quantitatively, we find the following fractional precision, normalized to SNR, for the QNM frequencies, in the event GW190521-like at $z=0.51$ in ET, 
\begin{align}   &\frac{\sigma_{\omega_{R,220}}}{ \omega_{R,220}}=0.094\% ; \quad \frac{\sigma_{\omega_{I, 220}}}{ \omega_{I,220}}=1.2\% , \\
     &\frac{\sigma_{\omega_{R, 440}}}{ \omega_{R, 440}}=0.53\% ; \quad \frac{\sigma_{\omega_{I, 440}}}{ \omega_{I, 440}}=15\% .\label{Dw_IMBH2440}
\end{align}
Comparing to the results of the previous section, we again find an improvement of less than $0.1\%$ in the (220) mode, and a factor of roughly 2 in the (440) mode. 

Interestingly, we now find the fractional errors in the QNM frequencies of the (330) mode to be larger than those of the (440) mode (compare to Eq.\ (\ref{Dw_IMBH330})). This is contrary to previous studies of linear QNMs, where the (330) mode is typically better measured than the (440) mode. We thus conclude that, due to the inclusion of the QQNM, the (440) mode becomes the most promising linear mode in future spectroscopic tests of gravity with such high-spinning  BBH events.

In Fig.\ \ref{fig:0507ADep} we compare the results on the precision of the QNM amplitudes, in the scenario of independent and dependent QQNM. Quantitatively, we obtain the following fractional precision:
\begin{align}
   &  \frac{\sigma_{|A_{220}|}}{|A_{220}|}=0.85\%   ; \quad  \frac{\sigma_{|A_{440}|}}{|A_{440}|}=10\% ,
\end{align}
which corresponds to an improvement less than $0.1\%$ in the (220) mode, and over a factor of two in the (440) mode (compare to Eqs.\ (\ref{ET_DA23})-(\ref{ET_DA44})). In the event at $z=0.51$ we obtain that the fractional amplitude precision is slightly better for the (440) than the (330) (compare to Eq.\ (\ref{ET_DA23_2})).

For this reason, in this case the maximum redshift satisfying the resolvability and detectability criteria is now determined by the precision of the (330) mode. We find that in order to satisfy the detectability and resolvability criteria for all three modes (220), (330), (440)
we require $z = 36$ for ET and $z = 24$ for CE. This is about one order of magnitude increase compared to the maximum redshifts obtained in the previous section when the QQNM is assumed to be independent. This happens because as the redshift increases not only the luminosity distance increases (which makes the GW amplitude decay linearly) but also the frequency is redshifted (which makes the ringdown signal lower its frequency and lie in a region where the detector noise is lower, in this case). Indeed, we find that at $z=36$ the ringdown SNR of the modes in the $(\ell,|m|)=(4,4)$ angular harmonic are only 4-5 times lower than at $z=0.51$. Since the maximum redshift is so high, we then expect all astrophysical intermediate-mass second-generation BBHs to have these three ringdown modes confidently detected, thanks to the improvements made by the QQNM.

We again provide the final fractional errors normalized to SNR, so that they can be used for estimating errors in similar binaries with different masses and redshifts: 

\begin{align}
    &\frac{\sigma_{\omega_{R,220}}}{ \omega_{R,220}}=0.15\% \left(\frac{100}{\rho_{22}}\right); \quad \frac{\sigma_{\omega_{I, 220}}}{ \omega_{I,220}}=2.0\%\left(\frac{100}{\rho_{22}}\right) , \\
     &\frac{\sigma_{\omega_{R, 440}}}{ \omega_{R, 440}}=2.0\% \left(\frac{10}{\rho_{44}}\right); \quad \frac{\sigma_{\omega_{I, 440}}}{ \omega_{I, 440}}=54\% \left(\frac{10}{\rho_{44}}\right),
\end{align}
and
\begin{align}
   &  \frac{\sigma_{|A_{220}|}}{|A_{220}|}=1.4\% \left(\frac{100}{\rho_{22}}\right)  ; \quad  \frac{\sigma_{|A_{440}|}}{|A_{440}|}= 38\% \left(\frac{10}{\rho_{44}}\right).
\end{align}
The results are again very similar to the ones obtained for the stellar-mas BBH, with all results changing by less than a factor of unity.

\section{Conclusions}\label{sec:conclusions}

We have performed Fisher forecasts on the ringdown signal of nearly equal-mass quasi-circular binary black hole mergers, for XG GW detectors, namely CE and ET. The ringdown signal in our model contains the three expected dominant linear QNMs, labeled by $(\ell m n)=$ (220), (330) and (440), as well as the expected dominant quadratic QNM coming from the self-interaction of the linear (220) mode.

We studied two representative binary systems: (i) stellar-mass BHs with low spin, describing a first-generation BBH merger, (ii) and intermediate-mass BHs with high spin, describing second-generation BBH mergers. We first confirm that the latter source leads to relative larger QQNM amplitudes, confirming results obtained in the past literature. 

For each one of the two types of binaries, we also analyze two different Fisher forecast scenarios: (a) the QQNM is considered to be an independent ringdown mode, and thus the Fisher forecast includes a total of four independent QNMs to be measured; (b) the QQNM is considered to be dependent on the QNM parameters of the linear (220) mode (based on GR), and thus the Fisher forecast includes a total of three independent linear QNMs to be measured. While Scenario (a) was analyzed in \cite{Yi:2024elj} (albeit in terms of BBH populations), to our knowledge this is the first time Scenario (b) is considered. When performing Fisher forecasts on each scenario, we impose resolvability and detectability constraints in order to quantitatively assess if all independent QNMs in the ringdown model will be detected confidently, and simultaneously, for performing future spectroscopic tests of gravity.

In Scenario (a) we generally find that that QQNM always has higher individual SNR than the (330) and (440) linear QNMs. Nonetheless, the QQNM is highly correlated with the (440) mode, and hence their parameter precision gets degraded. For this reason, we find the QQNM to be typically measured worse than the (330) mode but better than the (440) mode. For first-generation BBHs, all four QNMs are expected to pass the resolvability and detectability criteria when $z<0.045$ in ET, and $z<0.031$ in CE. For second-generation BBHs, the same maximum redshifts are found to be $z<0.51$ in ET, and $z<0.31$ in CE. We emphasize that these redshifts are driven by the amplitude of the (440) mode, which is the worst measured parameter of all. Therefore, these maximum redshifts are expected to be the same as those one would obtain in a purely linear theory (i.e.\ removing the QQNM from the ringdown model). For the binary masses considered here, for such limiting redshifts the QQNM parameters would be measured with percent-level precision in its real ($1-2\%$) and imaginary ($6-7\%$) frequency components, while its amplitude would be measured with $20-30\%$ precision. 

In Scenario (b) we generally find an improvement of a factor of two in the precision of the (440) QNM parameters, when compared to the results from Scenario (a). While the improvement on the (220) parameters is below $1\%$ and thus negligible. As a result of the improvement on the (440) mode, for first-generation BBHs, all three QNMs are expected to pass the resolvability and detectability criteria when $z<0.60$ in ET, and $z<0.33$ in CE. For second-generation BBHs, the same maximum redshifts are found to be $z<36$ in ET, and $z<20$ in CE. For this latter type of events, we find that the (440) mode is measured with better precision than the (330) mode, contrary to the results from Scenario (a) and those typically expected in a purely linear ringdown model. The redshift obtained is so high, that virtually all second-generation BBH mergers are expected to be excellent sources for performing consistency spectroscopic tests of GR with three ringdown QNMs. In order to compare to Scenario (a), we quote the parameter precision expected at the limiting redshifts obtained in Scenario (a).
In that case, both the (330) and (440) modes would have sub-percent precision on their real frequency components ($0.6-0.94\%$), $14-20\%$ precision on their imaginary frequency components, and $6.3-18\%$ precision on their amplitudes. 

We thus conclude that using the theoretical predictions from GR on  quadratic QNMs will lead to a one order of magnitude increase on the  maximum redshift of the BBH source for which multiple ringdown QNMs can be confidently detected simultaneously, and thus a large increase on the annual number of sources for which a consistency test of GR could be performed. A detailed analysis involving the expected BBH populations, similar to  \cite{Yi:2024elj}, could be performed for Scenario (b) to quantify the exact increase in source number.

Finally, we note that the results presented here are estimations based on idealized Fisher forecasts, as well as for specific noise models and detector designs. In the future, it would be useful to quantify how these results are affected when some of these assumptions are changed. Furthermore, by performing parameter estimations with a Bayesian analysis, one  could re-analyze Scenario (b), incorporating  the current uncertainty on the exact factor relating the QQNM amplitude to its parent modes for spinning black holes (c.f.\ Eq.\ (\ref{Qamp})), through the use of narrow priors on the QQNM amplitude . 

\section{ACKNOWLEDGMENTS}

We thank M.\ Isi and V.\ Baibhav for useful discussions during this project. 
The work of TA is supported in part by the Ministry of Science and Innovation (EUR2020-112157, PID2021-125485NB-C22, CEX2019-000918-M funded by MCIN/AEI/10.13039/501100011033), and by AGAUR (SGR-2021-01069).
ML was supported by the Innovative Theory Cosmology fellowship at Columbia University for partial time of this project. 
The work of JRV is supported, as part of his PhD Thesis, by the Formación de Profesorado Universitario Scholarship (FPU 20/04130) from Ministerio de Universidades (Spanish Government). LH acknowledges support from the DOE DE-SC011941 and a Simons Fellowship in Theoretical Physics.

\appendix

\section{SNR}\label{app:snr}

In frequency domain, the signal has the following expression \cite{Berti:2005ys}:
\begin{equation}\label{hf2}
      |\tilde{h}|^2 = \sum_{\ell m n n'} \frac{G^2M_z^2}{2c^4d_L^2} \left[ \frac{1}{10\pi}A^*_{\ell m n}A_{\ell m n'}\left(b_{\ell m n +}b_{\ell m n' +}
    + b_{\ell m n -}b_{\ell m n' -}\right) \right]   ,
\end{equation}
where we have already averaged over sky angles, using Eqs.\ (\ref{Faverage}), as well as over the binary angles $(\iota,\beta)$ using Eq.\ (\ref{Yaverage}). We emphasize that Eq.\ (\ref{hf2}) applies to linear and quadratic QNMs, since the QQNM considered here has the same role as an overtone in the $(4\pm 4)$ angular harmonics. For this reason, and to simplify notation, we assume $n$ can take the value $Q$ as well. 

In Eq.\ (\ref{hf2}) we have defined:
\begin{equation}\label{eq:bpm}
b_{\ell m n \pm}(2\pi f) \equiv \frac{-\omega^I_{\ell m n}}{(\omega^I_{\ell m n})^2 + (2\pi f \pm \omega^R_{\ell m n} )^2}.
\end{equation}
The total SNR can be calculated replacing Eq.\ (\ref{hf2})  into Eq.\ (\ref{Overlap_eqn}). In addition, for a single linear or quadratic QNMs in the $(\ell, |m|)$ angular harmonic, the SNR will be given by:
\begin{equation}\label{SNR1}
   \rho^2_{\ell m n}=\frac{4G^2M_z^2}{c^4d_L^2} \frac{|A_{\ell m n}|^2}{10\pi}\int_0^\infty df \frac{[b_{\ell m n+}^2(2\pi f)+b_{\ell m n-}^2(2\pi f)]}{S_n(f)},
\end{equation}
where we have already summed over positive and corresponding negative $m$ values.

\section{Fisher Matrix: Independent QQNM}\label{app:fisher_indep}
For each QNM, we will have four parameters $\{ |A_{\ell m n}|, \phi_{\ell m n}, \omega^R_{\ell m n}, \omega^I_{\ell m n}\}$, where $\phi_{\ell m n}$ is the phase of $A_{\ell m n}$. For each ringdown parameter $x$, we then calculate $\partial \tilde{h}/\partial x$ in frequency domain.

For any pair of QNMS with labels ($\ell m n$) and ($\ell m n'$), we calculate the fisher matrix, which will have the following components:
\begin{equation}
    \Gamma_{ij} = \frac{1}{10 \pi}  \frac{G^2M_z^2}{c^4d_L^2} \int_{0}^\infty df \;\frac{\gamma_{x_ix_j}(f)}{S_n(f)}
\end{equation}
where the explicit expressions for $\gamma_{x_ix_j}$ are:
\begin{align}
    & \gamma_{A_{\ell mn}A_{\ell m n'}}= 4\cos{\Delta_{\ell mnn'}}
  b_{\ell m n i}b_{\ell m n' i} \label{g_aa}\\
     & \gamma_{\phi_{\ell mn}\phi_{\ell m n'}}=  4|A_{\ell mn}||A_{\ell m n'}|\cos{\Delta_{\ell mnn'}}
    b_{\ell m n i}b_{\ell m n' i}\\
    & \gamma_{\omega_{P'\ell m n'}\omega_{P\ell mn}}=    4|A_{\ell m n'}||A_{\ell m n}|\cos{\Delta_{\ell mnn'}}
   b_{\ell m n i,P}b_{\ell m n' i,P'}
\end{align}

\begin{align}
   & \gamma_{A_{\ell mn'}\omega_{P\ell mn}}=    4|A_{\ell m n}|\cos{\Delta_{\ell mnn'}}
   b_{\ell m n i,P}b_{\ell m n' i} \label{g_aw}\\
     & \gamma_{\phi_{\ell mn}A_{\ell mn'}}=-  4|A_{\ell m n}|\sin{\Delta_{\ell mnn'}}
   b_{\ell m n i}b_{\ell m n' i}\\
    &\gamma_{\phi_{\ell mn}\omega_{P\ell mn'}}=-   4|A_{\ell m n}||A_{\ell m n'}|\sin{\Delta_{\ell mnn'}}
   b_{\ell m n i}b_{\ell m n' i,P}.
\end{align}
Here, we have used a compact notation, where $i$ can take the values $i=\{+,-\}$ and repeated $i$ indices refer to contracted sums:  e.g.\ $b_i b_i = b_+ b_+ + b_- b_-$. In addition, we have introduced the derivative terms $b_{\ell m n, P}$ where $P,P'\in \{R,I\}$ indicates derivatives of $b_{\pm}(2\pi f)$ (c.f.\ Eq.\ (\ref{eq:bpm})) with respect to $\omega_{\ell m n R}$ or $\omega_{\ell m n I}$. Omitting $(\ell m n)$ indices, the explicit expressions are:
\begin{align}
    & b_{\pm ,R}(\omega)=\frac{2(\omega_R\pm \omega)}{\omega_I}b^2_\pm (\omega), \\
    & b_{\pm ,I}=b_\pm (\omega)\left(\frac{1}{\omega_I}+2b_\pm (\omega)\right).
\end{align}

In addition, we have introduced the relative mode phase shift
\begin{equation}\label{Gmatrix}
    \Delta_{lmnn'}\equiv \phi_{lmn}-\phi_{lmn'}.
\end{equation}

Any other components, such as those for two modes with different $\ell m$, will be vanishing. This means that the total 16x16 fisher matrix can be expressed in blocks as:
\begin{equation}
   \Gamma= \begin{pmatrix}
   \Gamma^{22} & \mathbf{0} & \mathbf{0} \\ \mathbf{0} & \Gamma^{33} & \mathbf{0}\\
   \mathbf{0} & \mathbf{0} & \Gamma^{44}
    \end{pmatrix},
\end{equation}
where $\mathbf{0}$ stand for 4x4 vanishing matrices. Here, $\Gamma^{22}$ and $\Gamma^{33}$ are 4x4 matrices whose only non-vanishing components are those in Eqs.\ (\ref{g_aa})-(\ref{g_aw}), for the (220) and (330) modes respectively. Instead, $ \Gamma^{44}$ is a 8x8 matrix which includes the mixing between the $(440)$ and $(44Q)$ modes. 

\section{Fisher Matrix: Dependent QQNM}\label{app:fisher_dep}
In this scenario we assume the quadratic QNM is directly related to the linear 220 mode (both in amplitude and frequency):
\begin{align}
    &A_{44Q}=\alpha e^{i\delta }A^2_{220},\\
    & \omega_{44Q}=2\omega_{220},
\end{align}
with $\alpha$ real.
In this section, we keep $\alpha$ and $\delta$ generic, but in the results presented in the main sections we use $\alpha=0.154$ and $\delta=-0.068$rad, according to Eq.\ (\ref{Qamp}).

The total fisher matrix will again be expressed in blocks but differently to the previous scenario:
\begin{equation}
   \Gamma= \begin{pmatrix}
   \Gamma^{22} & \mathbf{0} & \Gamma^{22x 44} \\ \mathbf{0} & \Gamma^{33} & \mathbf{0}\\
   \Gamma^{22x 44} & \mathbf{0} & \Gamma^{44}
    \end{pmatrix},
\end{equation}
The matrix $\Gamma^{33}$ will be the same as for the `independent QQNM' case (and thus we use the expressions in Appendix \ref{app:fisher_indep}). Nonetheless, the individual elements that will appear in $\Gamma^{22}$ and $\Gamma^{44}$ will be modified, and in addition we expect to have mixing terms between the (220) and (440) modes encoded in the $\Gamma^{22x44}$ matrix. This mixing occurs because the (44Q) harmonic depends on the (220) QNM parameters, and since it appears in the $(\ell m=44)$ angular harmonic it mixes with the (440) QNM.
Here, each block $\Gamma^{22}$, $\Gamma^{44}$, and $\Gamma^{22x44}$ has 4x4 dimensions.

The components of $\Gamma^{22}$ will be given by:
\begin{align}
 \gamma_{A_{220}A_{220}} &= 4 b_{220 i}b_{220 i}  + 16\alpha^2 |A_{220}|^{2} b_{44Q i}b_{44Q i} \\
 \gamma_{\phi_{220}\phi_{220}} &=   4|A_{220}|^2  b_{220 i} b_{220 i} + 16\alpha^2 |A_{220}|^{4} b_{44Q i} b_{44Q i}
 \end{align}
\begin{align}
\gamma_{\omega_{P220}\omega_{P'220}} &=  4|A_{220}|^2 b_{220 i,P} b_{220 i,P'}  \nonumber \\
& + 16\alpha^2|A_{220}|^4 b_{44Q i,P} b_{44Qi,P'} \\
\gamma_{A_{220}\omega_{P220}} & = 4|A_{220}| b_{220 i,P} b_{220 i}  \nonumber \\
  &+  16\alpha^2 |A_{220}|^{3} b_{44Q i,P} b_{44Q i}
\end{align}
with $\gamma_{\phi_{220}A_{220}}= \gamma_{\phi_{220}\omega_{R220}}=\gamma_{\phi_{220}\omega_{I220}} = 0$.

The components of the mixing matrix $\Gamma^{22x44}$ will be given by:
\begin{align}
& \gamma_{A_{220}A_{440}} =  8\alpha|A_{220}|\cos \Delta \; b_{44Q i} b_{440 i} \\
& \gamma_{A_{220}\phi_{440}} =  8\alpha|A_{220}||A_{440}| \sin \Delta \; b_{44Q i} b_{440 i}\\
& \gamma_{A_{220}\omega_{P440}} =  8\alpha|A_{220}||A_{440}|\cos \Delta \;b_{44Q i} b_{440 i,P}\\
 &   \gamma_{\phi_{220}\phi_{440}} =  8 \alpha|A_{220}|^2|A_{440}|\cos \Delta \; b_{44Q i} b_{440 i}\\
&\gamma_{\phi_{220}A_{440}} =    -8\alpha|A_{220}|^2\sin \Delta \; b_{44Q i} b_{440 i}  \\
&\gamma_{\phi_{220}\omega_{P440}} = - 8\alpha|A_{220}|^2|A_{440}| \sin \Delta \; b_{44Q i} b_{440 i,P}\\
&\gamma_{\omega_{P440}\omega_{P'220}} =  8\alpha|A_{440}||A_{220}|^2\cos \Delta \; b_{44Q i,P'} b_{440 i,P}\\
&\gamma_{\omega_{P220}A_{440}} = 8\alpha|A_{220}|^2\cos \Delta \; b_{44Q i,P} b_{440 i}\\
&\gamma_{\omega_{P220}\phi_{440}} = 8\alpha|A_{220}|^2|A_{440}|\sin \Delta \; b_{44Q i,P} b_{440 i} ,
\end{align}
where $\Delta  \equiv  2 \phi_{220} + \delta  - \phi_{440}$.
Here, we use the same notation introduced in Appendix \ref{app:fisher_indep}.

\bibliographystyle{apsrev4-1}
\bibliography{References.bib}

\end{document}